\documentclass[aps, prl, twocolumn, superscriptaddress]{revtex4-1}
\usepackage{graphicx}
\usepackage{dcolumn}
\usepackage{natbib}
\usepackage{multibib}
\usepackage{amssymb}
\usepackage{amsmath}

\begin{document}
\title{First Principles Prediction of Topological Phases in Thin Films of Pyrochlore Iridates}
\author{Xiang Hu} 
\email{phyxiang@gmail.com}
\affiliation{Department of Physics, The University of Texas at Austin, Austin, Texas 78712}
\author{Zhicheng Zhong}
\affiliation{Institute of Solid State Physics, Vienna University of Technology, A-1040 Vienna, Austria}
\author{Gregory A. Fiete}
\affiliation{Department of Physics, The University of Texas at Austin, Austin, Texas 78712}
\date{\today}

\begin{abstract}
Topological phases have attracted much interest in recent years\cite{Moore:nat10,Hasan:rmp10,Qi:rmp11}.  While there are a number of three-dimensional materials exhibiting topological properties\cite{Ando:jpsj13}, there are relatively few two-dimensional examples \cite{Konig:sci07,Knez:prl11,Chang:sci13} aside from the well-known quantum Hall systems.  Here we make materials-specific predictions for topological phases using density functional theory combined with Hartree-Fock theory that includes the full orbital structure of the relevant iridum $d$-orbitals and the strong but finite spin-orbit coupling strength.  We find Y$_2$Ir$_2$O$_7$ bilayer and trilayer films grown along the [111] direction can support topological metallic phases\cite{Bergman:prb10,Pan_top_met_14} with a direct gap of up to 0.02 eV, which could potentially bring transition metal oxides to the fore as a new class of topological materials with potential applications in oxide electronics \cite{Hwang:nm12}
\end{abstract}

\maketitle

While there have been numerous experimental discoveries of three-dimensional topological insulators (TI), there remains the persistent problems of high bulk conductivity and surface state properties that can change significantly over the course of a few hours or days\cite{Ando:jpsj13}.  Therefore, it is desirable to seek out new material classes that might support topological phases.   Transition metal oxides (TMO) with heavy $4d$ or $5d$ transition metal ions, such as iridium, have drawn considerable experimental and theoretical interest in this regard\cite{Krempa:arcm14}.  Particularly on the theoretical side, the pyrochlore iridates have motivated a number of studies predicting novel topological phases\cite{Pesin:np10,Kargarian:prb11,Kargarian:prl13,Maciejko:prl14,Go:prl12,Wan:prb11}.
 
In addition to the bulk TMO as candidates for supporting three-dimensional topological phases, several thin-film TMO systems have also been suggested\cite{Ruegg11_2,Yang:prb11a,Ruegg:prb12,Xiao:nc11,Okamoto:prb14,Yang:prl14,Okamoto:prl13,Doennig:prb14,Lado:prb13,Liang:njp13,Wang14,Hu:prb12}.    In the thin-film  TMO, both the time-reversal invariant $Z_2$ TI\cite{Ruegg11_2,Yang:prb11a,Xiao:nc11,Hu:prb12,Lado:prb13,Liang:njp13} and the time-reversal symmetry broken quantum anomalous Hall or Chern insulator (CI) states\cite{Ruegg11_2,Yang:prb11a,Ruegg:prb12,Xiao:nc11,Yang:prl14,Okamoto:prl13,Doennig:prb14,Wang14,Hu:prb12}  (which exhibit a quantized Hall conductance in zero applied magnetic field) have been predicted.   Both the two-dimensional TI and CI states rely on spin-orbit coupling.  In the context of thin-film TMO, the spin-orbit coupling may be ``dynamically generated" through interactions for light transition metals\cite{Ruegg11_2,Yang:prb11a},  or may be intrinsic for oxides with heavy transition metal ions\cite{Hu:prb12}.  In this work, we use a combination of first-principles calculations and Hartree-Fock theory to show bilayer and trilayer films of Y$_2$Ir$_2$O$_7$ grown along  the [111] direction can support topological metallic phases\cite{Bergman:prb10,Pan_top_met_14} with a direct gap of up to 0.02 eV.   Under the right perturbations (which can include substrate strain and charge density wave order), these phases can be driven to their insulating topological counterparts.

We study a sandwich structure consisting of a thin (few atomic layers) Y$_2$Ir$_2$O$_7$ film between the non-magnetic, large-gap band insulator Y$_2$Hf$_2$O$_7$, as shown in Fig.\ref{fig:films} (a).  We find quantitatively similar results if   La$_2$Ir$_2$O$_7$ replaces Y$_2$Ir$_2$O$_7$.   The ``capping layers" of Y$_2$Hf$_2$O$_7$ serve mainly to stabilize the thin Y$_2$Ir$_2$O$_7$ films, but also strain the films and modify their band structure compared to the unstrained case.  In our study, we used first-principles calculations (see Methods) to determine the band structure of the system, and included different thicknesses of Y$_2$Hf$_2$O$_7$ in fully relaxed structures (for a fixed Y$_2$Ir$_2$O$_7$ film, either bilayer or trilayer)  to assess the magnitude of the strain effects.  Our calculations show the strain from the lattice mismatch (1.9\%)  between Y$_2$Ir$_2$O$_7$ and Y$_2$Hf$_2$O$_7$ is not a large effect in this system (possibly because of the spatially extended Ir $d$-orbitals), and therefore we do not believe the choice of  Y$_2$Hf$_2$O$_7$ is integral to the physics of the Y$_2$Ir$_2$O$_7$ films. Other choices of wide-gap band insulator ``capping layers" could be made to investigate strain effects in more detail.  Using the DFT results, we performed a Wannier projection and treated the large $d$-orbital interactions separately within the full $t_{2g}$ subspace using unrestricted Hartree-Fock theory.   Our main result is the phase diagram shown in Fig.\ref{fig:phase_diagram}, which shows a topological metallic phase for small interactions in the bilayer, and a topological Chern metal in the trilayer for a range of interactions around 0.5 eV.  

\begin{figure}
\centering
\hspace{-31.0mm}
\includegraphics[width=0.60\linewidth]{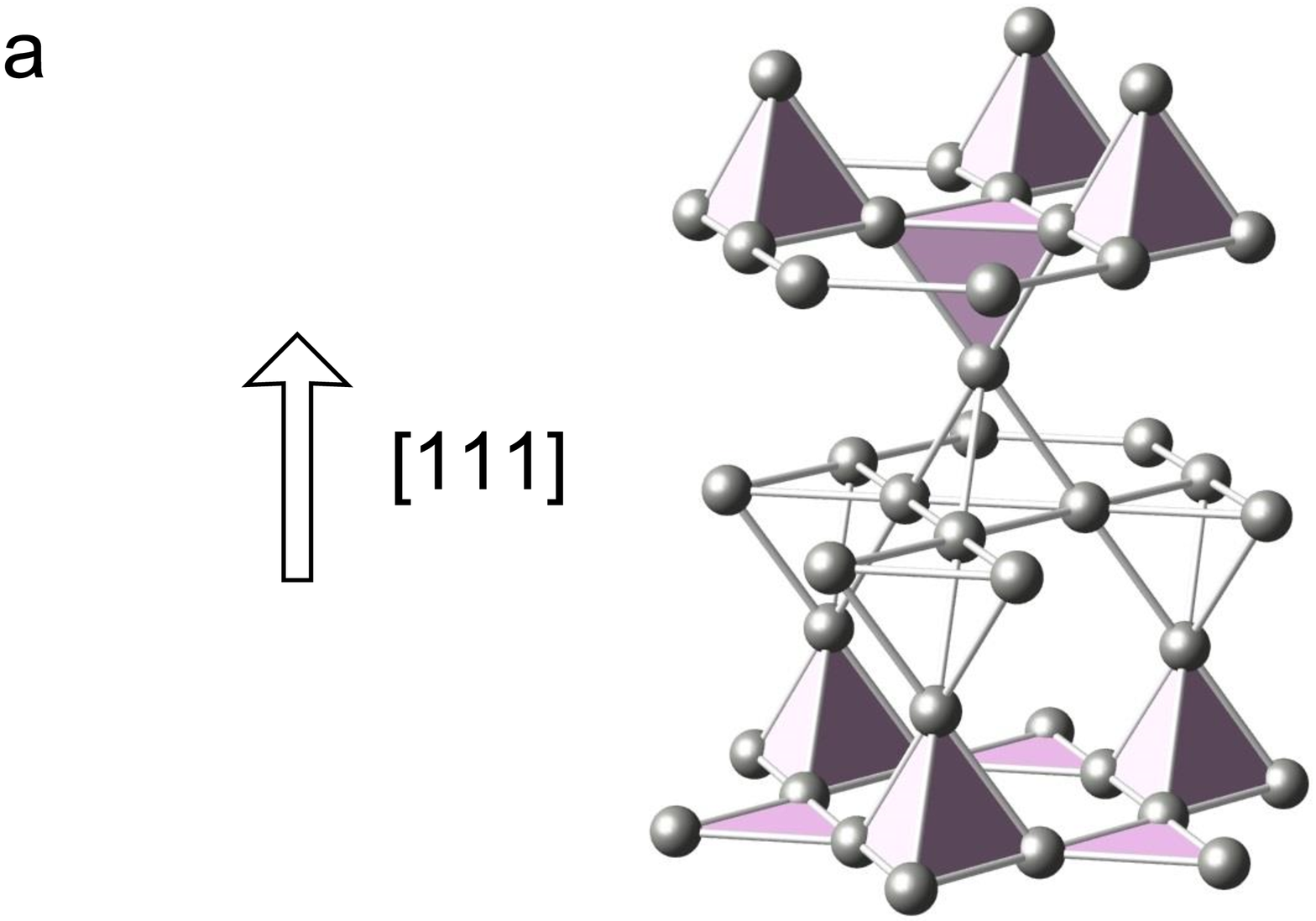}
\includegraphics[width=1.00\linewidth]{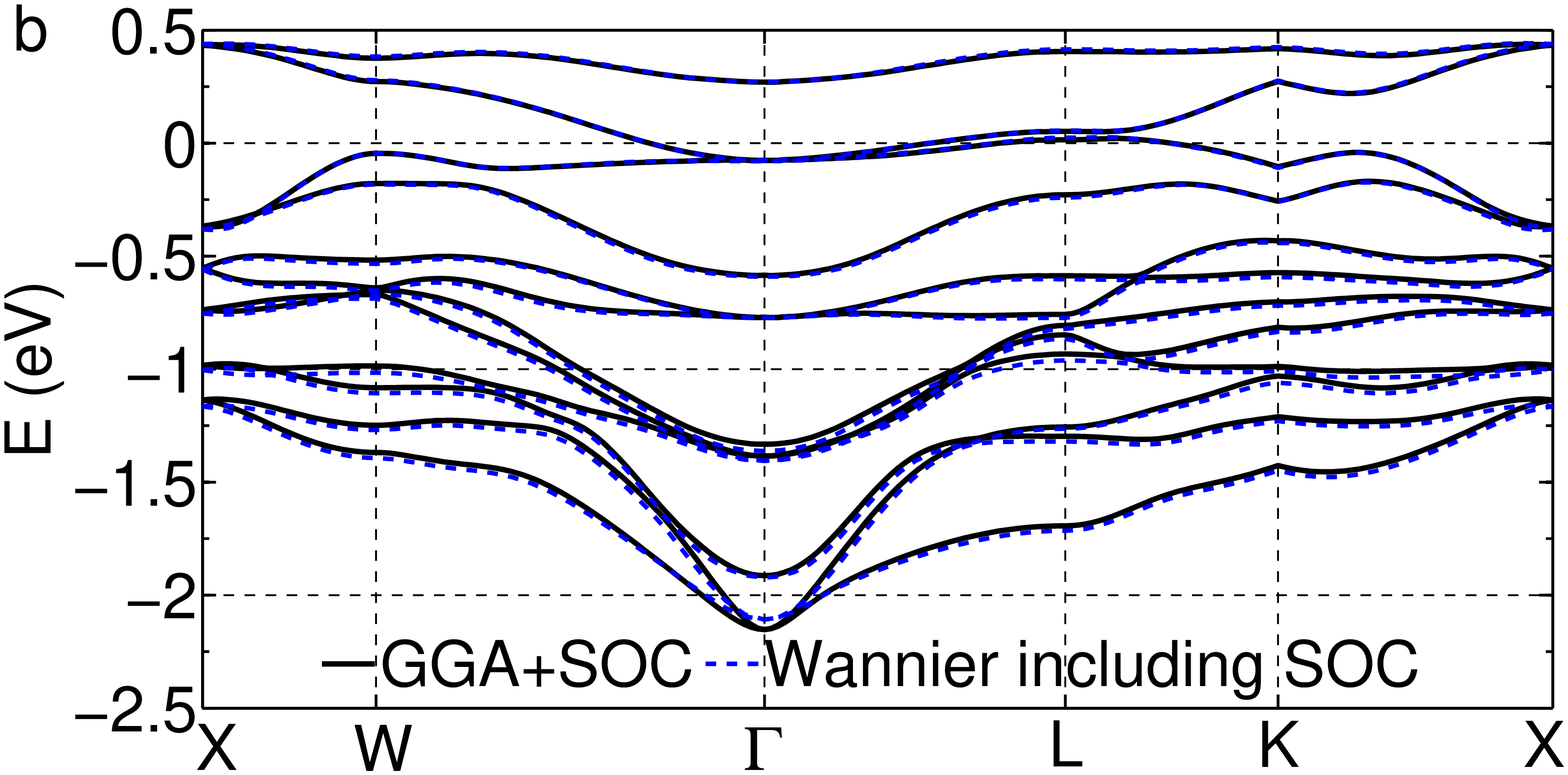}
\caption{(a) The bulk crystal structure of Y$_2$Ir$_2$O$_7$. Only the Ir$^{4+}$ ions
are displayed. The bilayer film structure is shaded purple/lavender on the bottom and the 
triangular-kagome-triangular (TKT) film structure shaded on the top. (b) The bulk 
band structure of Y$_2$Ir$_2$O$_7$ determined using the generalized gradient approximation (GGA) plus spin-orbit coupling (SOC) [solid black line] and the Wannier projection [dotted blue line].   \label{fig:bulk_bands}
}
\end{figure}

To ground our study, we initially performed first principles electronic structure calculations (see Methods) on the {\em bulk} Y$_2$Ir$_2$O$_7$ system, for which prior results are available\cite{Wan:prb11}.   A portion of the pyrochlore lattice of iridium ions is shown in Fig.\ref{fig:bulk_bands} (a), and the electronic band structure is shown in Fig.\ref{fig:bulk_bands} (b), along with the results from a model fit with $t_{2g}$-like Wannier projected states (dashed blue line) with the spin-orbit coupling turned on. Our first-principles (DFT) calculations were carried out in the generalized gradient approximation (GGA) of the exchange-correlation potential with a $10\times10\times10$ k-point grid.  The computations were
performed in two different ways: (i) By using the all-electron full-potential 
augmented plane-wave method in the \texttt{Wien2k} implementation, and (ii) By using a 
norm-conserving pseudopotential in the \texttt{Quantum Espresso} implementation. The two different packages
obtain very close results. The spin-orbit coupling was 
included in the fully-relativistic schemes, and included non-zero spin-orbit coupling on all the atoms.
We took the experimental structure of bulk Y$_2$Ir$_2$O$_7$ with space group (227, Fd-3m). Each unit cell contains four equivalent Ir$^{4+}$ sites. Because the structure parameter $x=0.3356>x_c=5/16$ in this material, each
oxygen octahedral cage is subjected to a trigonal distortion, which  splits the $t_{2g}$ orbitals into $e'_g$ and $a_{1g}$ states.  The Wannier projections including spin-orbit coupling were carried out with a $8\times8\times8$ 
mesh size with an initial basis of the local $t_{2g}$ orbitals at each Ir$^{4+}$ site with the spin-quantization axis set to the global z-axis.  Evidently, the Wannier fit and the direct density functional theory (DFT) result are in fairly good agreement, including important details such as the degeneracies at the $\Gamma$ point for bands near the Fermi energy (set equal to zero).  In addition, one sees that the upper 4 bands (roughly, the total angular momentum $j=1/2$ states) and the lower 8 bands (roughly, the total angular momentum $j=3/2$ states) are {\em not} well separated from each other.   Because of the combination of time-reversal symmetry (we focused on non-magnetic solutions within DFT) and inversion symmetry, each band is doubly degenerate, for a total of 24 $t_{2g}$-type bands originating from the 4 iridium atoms in the pyrochlore unit cell.  (Strictly speaking, there is a small mixing of oxygen $p$-type orbitals into the bands shown as well.)  Overall, these results compare well with those published\cite{Wan:prb11}; the small differences can mostly be attributed to the additional $U$-term (not included in Fig.\ref{fig:bulk_bands} (b), but included in\cite{Wan:prb11}) which helps stabilize magnetic order.  We used the non-magnetic DFT result (without a ``$U$") as input for a Hartree-Fock calculation (details below) that includes interactions within the {\em full} $5d$ $t_{2g}$-orbital manifold, {\it i.e.}, {\em we did not} make a strong spin-orbit coupling ($j=1/2$) approximation and treat interactions and kinetic terms of the Hamiltonian within this restricted subspace.  Using the combination of DFT and Hartree-Fock theory for bulk Y$_2$Ir$_2$O$_7$, we find the same ``all-in/all-out" magnetic state reported earlier\cite{Wan:prb11} for a moderate Hubbard $U$ value (see Eq.\eqref{eq:H_int}) of 0.7-1.8 eV.  This establishes that our approach and methods can capture the important details obtained in earlier calculations for the bulk systems.

\begin{figure*}
\centering
\includegraphics[width=0.90\linewidth]{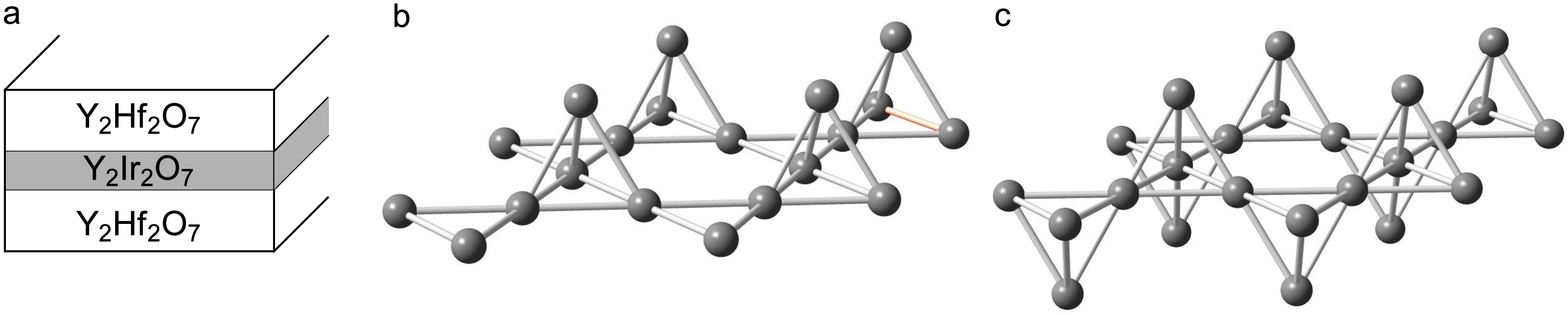}
\includegraphics[width=0.43\linewidth]{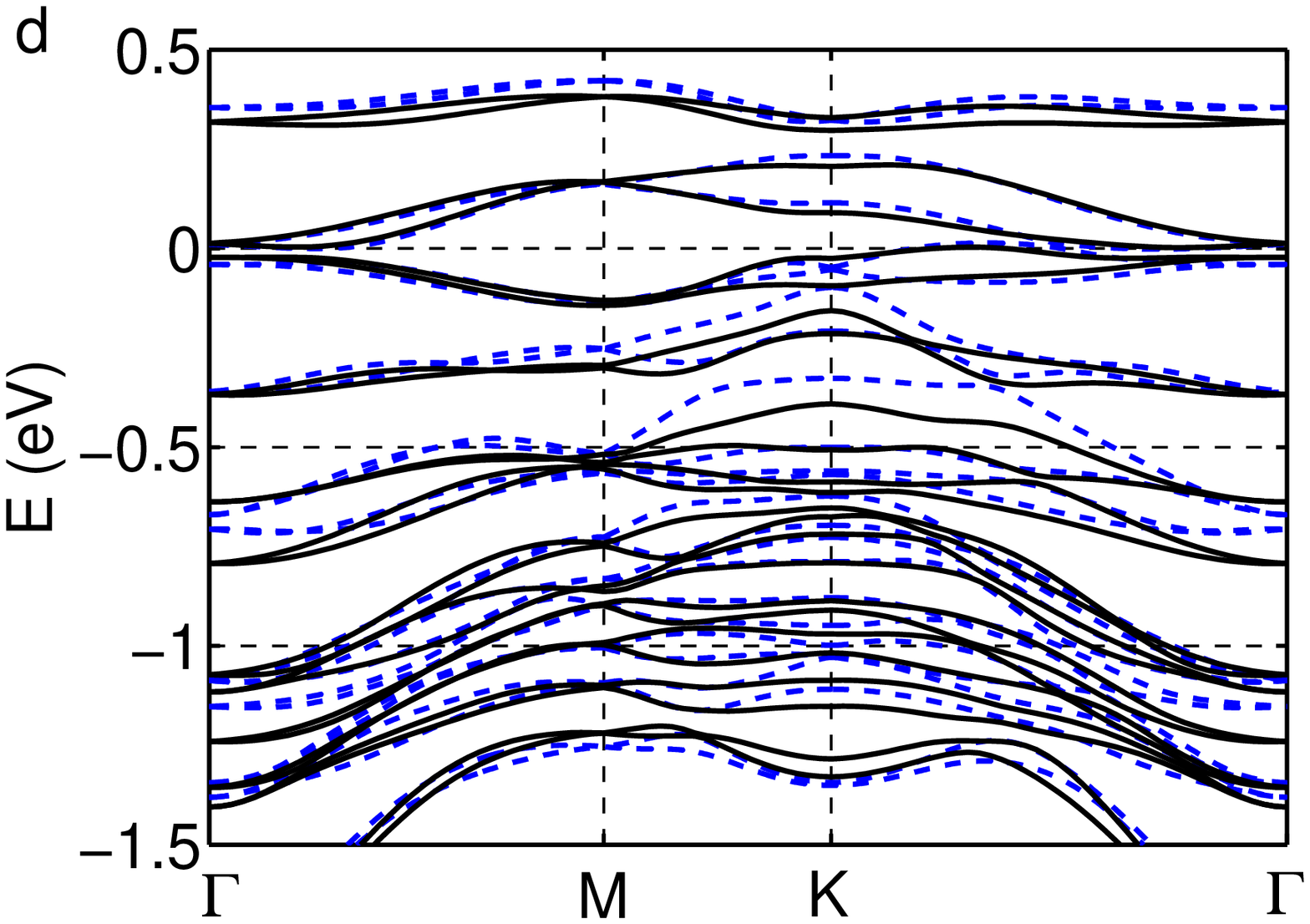}
\includegraphics[width=0.43\linewidth]{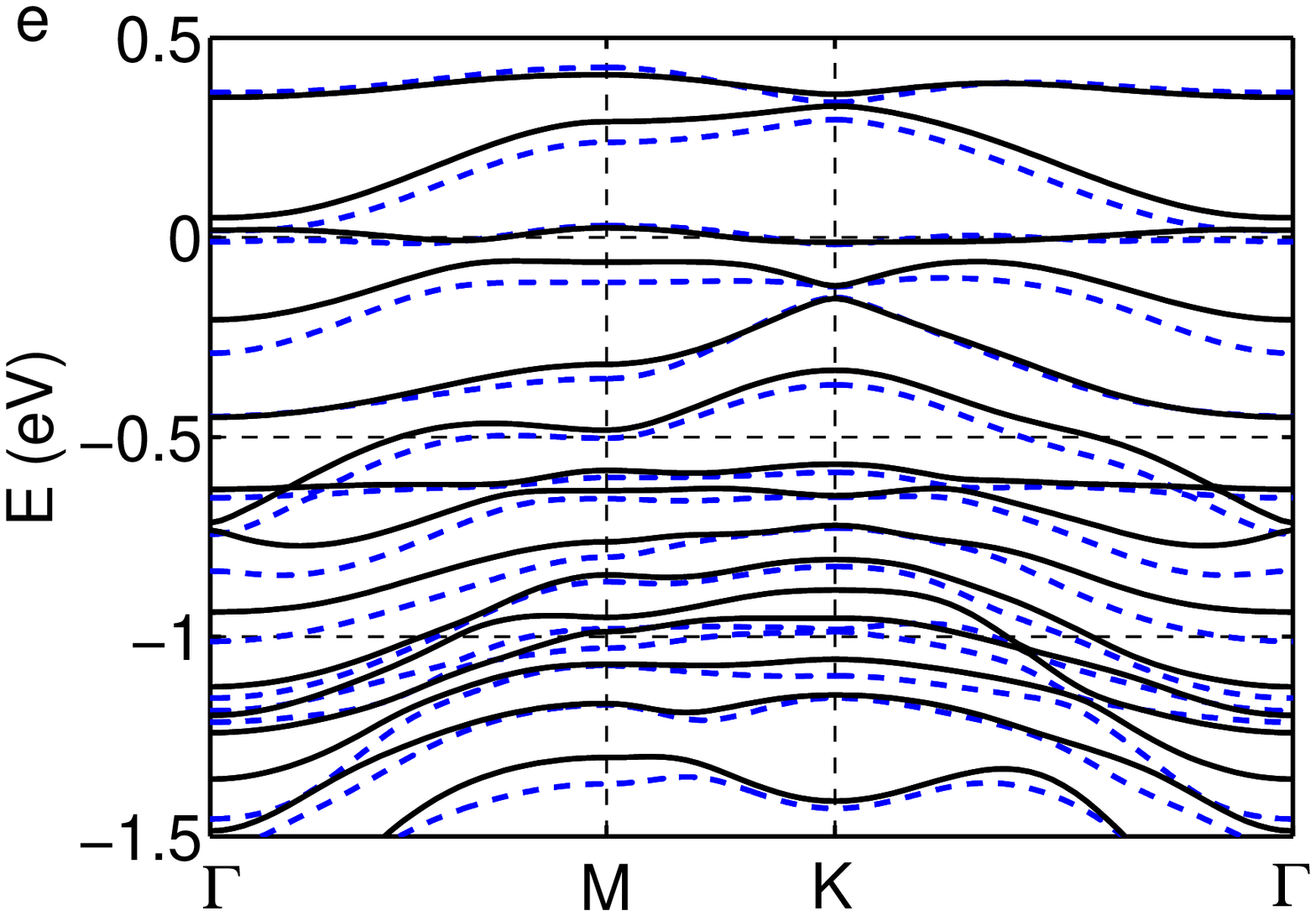}
\caption{Thin film structures and corresponding electron energy bands. (a) Schematic of the thin film ``sandwich" structure. (b) Bilayer thin film. (c) TKT thin film. Only the 
Ir$^{4+}$ ions are shown in (b) and (c). (d) Bilayer electronic band structure from a fully 
relaxed GGA+SOC calculation [solid black] with Wannier fit [dashed blue]. (e) TKT band structure from a fully relaxed 
GGA+SOC calculation [solid black] with Wannier fit [dashed blue].  The Wannier fits (obtained from bulk Wannier projection) are used in the Hartree-Fock calculation of the phase diagrams of the bilayer and trilayer.\label{fig:films}}
\end{figure*}

We now turn our attention to the thin-film systems.  The primary systems of interest are the [111]-grown bilayer (bottom of Fig.\ref{fig:bulk_bands} (a) and Fig.\ref{fig:films} (b)) and triangular-kagome-triangular (TKT) trilayer films (top of Fig.\ref{fig:bulk_bands} (a) and Fig.\ref{fig:films} (c)).  
The thin film DFT calculations were carried out in the (Y$_2$Ir$_2$O$_7$)$_2$/(Y$_2$Hf$_2$O$_7$)$_2$ and (Y$_2$Ir$_2$O$_7$)$_3$/(Y$_2$Hf$_2$O$_7$)$_3$ superlattices, whose structure was fully relaxed. 
The non-magnetic electronic band structure along with Wannier fits are shown in Fig.\ref{fig:films} (d) for the bilayer and in Fig.\ref{fig:films} (e) for the TKT trilayer, which indicates both systems are metallic.  Compared with the bulk electronic band structure shown in Fig.\ref{fig:bulk_bands} (b), the quality of the Wannier fit is less good for the films.  (See supplemental for details.)  We attribute this to the lower symmetry of the films compared to the bulk and the fact that the iridium $5d$-orbitals are rather extended.  As we show in the Supplemental information, a {\em tight-binding fit} to the bulk band structure that includes out to third neighbor hopping yields a rather poor fit to the DFT results.  As a result, the iridium $5d$-orbitals in the bilayer and trilayer films have a spatial extent that is at least comparable to the film thickness and therefore the $5d$-orbitals are influenced by the ``capping"  Y$_2$Hf$_2$O$_7$ layers and experience a local environment of reduced symmetry.  While this leads to a slightly less good Wannier fit than we obtained for the bulk, the bands and the density of states around the Fermi energy are reasonably close.  Therefore, we expect there to be little numerical difference in a Hartree-Fock calculation that makes use of the Wannier fit shown, and one that makes use of a fit that better approximates the DFT result. 
We note that since the bilayer breaks inversion symmetry, the bands in Fig.\ref{fig:films} (d) are non-degenerate, while the bands of the inversion symmetry-preserving TKT layer in Fig.\ref{fig:films} (e) are two-fold degenerate.  As we mentioned earlier, the larger spatial extent of the $5d$-orbitals compared to the $3d$-orbitals is expected to make TMO films with heavy transition metals less susceptible to interfacial strain--a quality that might enhance the robustness of theoretical predictions for topological phases in thin-film systems with heavier elements. 

After obtaining a {\em non-interacting} Hamiltonian in a local ($t_{2g}$-like Wannier) basis for the bilayer and TKT trilayer, 
\begin{equation}
\label{eq:H_0}
H_0=\sum_{i,j,\alpha,\beta}\tilde{t}_{i\alpha, j\beta}
c_{i\alpha}^{\dagger}c_{j\beta},
\end{equation}
we studied the full Hamiltonian $H=H_0+H_U$ where
\begin{equation}
\label{eq:H_int}
H_U=\frac{U}{2}\sum_i\left(\sum_{\alpha} \hat{n}_{i\alpha}-5\right)^2, 
\end{equation}
within Hartree-Fock theory.   In Eq.\eqref{eq:H_0} and Eq.\eqref{eq:H_int}, $i, j$ represent different sites, $\alpha, \beta$ indexes different orbitals (including spin indicies) in the $t_{2g}$ manifold, and $c_{i\alpha}^{\dagger}$ ($c_{i\alpha}$) is the creation (annihilation) operator of an electron on site $i$ with orbital $\alpha$.  The complex hopping amplitude $\tilde{t}_{i\alpha, j\beta}$ includes spin-orbit coupling, and are obtained from the Wannier fit to the DFT results.  The rotationally invariant (in orbital and spin space) Hubbard term, Eq.\eqref{eq:H_int}, follows the convention used in previous works\cite{Kargarian:prb11}, and $\hat{n}_{i\alpha}$ is the number operator for site $i$, orbital $\alpha$.  The rotationally invariant form allows the Hartree-Fock calculation to be performed in any basis, and allows us to extend most previous studies on the iridates by explicitly including the full $t_{2g}$ subspace and keeping the spin-orbit coupling finite, as opposed to working in a strong (infinite) spin-orbit coupling limit with only $j=1/2$ states in the Hilbert space.  The main approximation made in the form of the interaction given in Eq.\eqref{eq:H_int} is the neglect of the Hund's coupling.  Physically, this restricts our results to low-spin configurations on the iridium ions, which is consistent with prior work\cite{Wan:prb11,Zhang_Huale:prl13}.

The Hartree-Fock calculation is performed by decoupling the on-site interactions, Eq.\eqref{eq:H_int}, as
\begin{eqnarray}
\noindent
H_U&=&\frac U 2\sum_{i}\sum_{\alpha\beta,\alpha\neq\beta}
\left[2n_{i\alpha}\hat{n}_{i\beta}-n_{i\alpha}n_{i\beta} \right.\nonumber\\
&&\left. -2m^i_{\alpha\beta}\hat{m}^i_{\beta\alpha}+|m^i_{\alpha\beta}|^2\right],
\end{eqnarray}

where $\hat{n}_{i\alpha}=\hat{c}_{i\alpha}^{\dagger}\hat{c}_{i\alpha}$, and
$\hat{m}^i_{\alpha\beta}=\hat{c}_{i\alpha}^{\dagger}\hat{c}_{i\beta}$ with $\alpha \neq \beta$.  We perform an unrestricted Hartree-Fock calculation in which the filling in the unit cell is held fixed while the electron density on each site is allowed to vary.  The self-consistent calculation generates a local minimum of the total energy through the variation of the coupling constants $\langle \hat{n}_{i\alpha}\rangle$ and $\langle \hat{m}^i_{\alpha\beta}\rangle$, which are determined by the eigenstates below the Fermi level.  $Z_2$ and Chern invariants are calculated as described in the supplemental material.  The Hartree-Fock results for the [111]-grown bilayer and TKT trilayer  are shown in Fig.\ref{fig:phase_diagram}, where the strength of $U$ is varied.  Because $U$ is an effective interaction parameter within the $t_{2g}$ manifold, it is difficult to determine its precise value from experiment or theory.  Estimates in the literature range from 0.4-2.0 eV\cite{Krempa:arcm14}, so we include this range in our calculations.

Based on the band structures shown in Fig.\ref{fig:films}, the small-$U$ behavior is expected to be metallic, and we indeed find this is the case.  From Figs.\ref{fig:films} (d), (e) one can see that the band width of the states closest to the Fermi energy are on the order of 0.5 eV.  Thus, one expects possible transitions when $U$ is of the order of 0.5 eV, and one again finds that in both systems time-reversal symmetry in spontaneously broken ({\it i.e.}, magnetism sets in) for interactions of this rough magnitude.  By investigating the small $U$ metallic phases in both cases, we find that the metallic state in the bilayer is actually a topological metal\cite{Hasan:rmp10}: While the direct gap is finite throughout the Brillouin zone, the indirect gap is negative.  The finite direct gap allows one to compute the $Z_2$ invariant for the lowest 20 bands, and we find that it is topologically non-trivial.  Therefore, if one were able to deform the bands so that the direct gap remains open while the indirect gap is made positive, one would obtain a $Z_2$ topological insulator.  Because the change in energy of states near the Fermi energy required to do this is of the order of a few tens of meV, a judicious choice of substrate in experiment may in fact turn the bilayer system into a $Z_2$ topological insulator if interactions are sufficiently screened with a nearby metallic gate so that the effective $U$ value is reduced compared to the bulk value.

\begin{figure}
\includegraphics[width=1.00\linewidth]{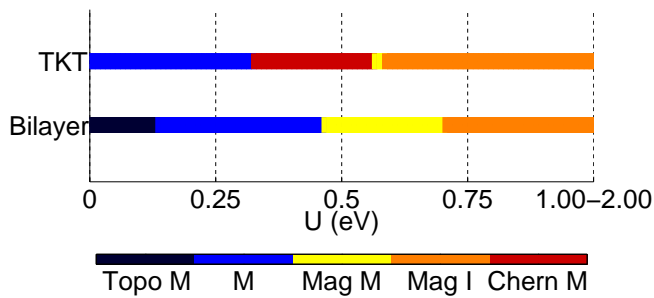}
\caption{Hartree-Fock phase diagram of the bilayer and TKT thin films.  For small $U$ the bilayer exhibits a topological metal phase with time-reversal symmetry preserved.  For a somewhat larger $U$, time-reversal symmetry is spontaneously broken for both the bilayer and trilayer, with a Chern metal appearing in the TKT film centered around $U\approx 0.5$ eV. If disorder in the system is not too strong, edge modes in the topological metal and Chern metal can behave qualitatively similar to their insulating counterparts. Top M=topological metal, M=metal, Mag M=magnetic metal,  Mag I=magnetic insulator, and Chern M = Chern Metal.
\label{fig:phase_diagram}}
\end{figure}

\begin{figure}
\centering
\includegraphics[width=0.87\linewidth]{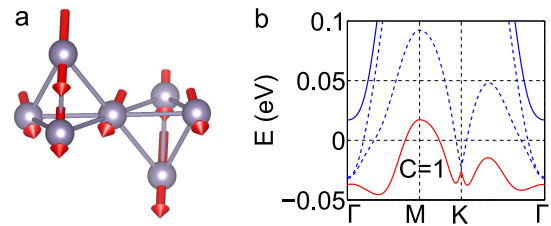}
\caption{The electronic band structure and local magnetic moments of the Chern metal phase in the TKT
thin film when $U=0.43$eV. (a) Orientation of the local magnetic moments. (b) The electronic band structure close to the Fermi energy, which we have set to zero. The red band possesses nonzero Chern
number.
\label{fig:moments}}
\end{figure}

A closer examination of the TKT trilayer also reveals a topological metallic phase, the Chern metal around $U\approx 0.5$ eV.   The situation in this case is qualitatively similar to the topological metal in the bilayer, except that the Chern metal has broken time-reversal symmetry:  The Chern metal has a finite direct gap throughout the Brillouin zone, but a negative indirect gap such that the lowest bands have a total Chern number of 1 so that a deformation of the Hamiltonian that keeps the direct gap open but makes the indirect gap positive would result in a Chern insulator.  In contrast to the results for the bulk system, we do not find an ``all-in/all-out" magnetic order, but instead a deformation of it (shown in Fig.\ref{fig:moments}) that has a net magnetic moment.

In summary, we have reported a combined first-principles and Hartree-Fock study of bilayer and trilayer films of Y$_2$Ir$_2$O$_7$.  Using realistic electronic band structure with a finite spin-orbit coupling included, we calculate the effect of interactions within the full $t_{2g}$ iridium $5d$-orbital subspace and find the thin films systems may support topological metallic phases.   These topological metallic phases could be converted into their insulating counterparts with the right substrate strain.  We hope this work will help encourage further experimental efforts in this direction and facilitate the discovery of topological phases in transition metal oxides.

We thank Liang Du, Craig Fennie, Priyamvada Jadaun, Mehdi Kargarian, Penghao Xiao, Chandrima Mitra, and Andreas Ruegg for helpful discussions.  Our work was generously funded by ARO Grant No. W911NF-09-10527, NSF Grant No. DMR-0955778, and DARPA grant No. D13AP00052. We thank The Texas Advanced Computing Center (TACC) at The University of Texas at Austin for providing the necessary computing resources. URL:http://www.tacc.utexas.edu.

%


\onecolumngrid
\newpage

\setcounter{equation}{0}
\setcounter{figure}{0}
\setcounter{table}{0}
\setcounter{page}{1}
\makeatletter
\renewcommand{\theequation}{S\arabic{equation}}
\renewcommand{\thefigure}{S\arabic{figure}}
\renewcommand{\bibnumfmt}[1]{[#1]}
\renewcommand{\citenumfont}[1]{#1}

\begin{center}
\textbf{\large Supplemental Materials for ``First Principles Prediction of 
Topological Phases in Thin Films of Pyrochlore Iridates"}
\end{center}

\section{I. Bulk DFT calculation for Y$_2$I\MakeLowercase{r}$_2$O$_7$}

The DFT calculation of bulk Y$_2$Ir$_2$O$_7$ was carried out with \texttt{Wien2k}  \cite{Blaha:tu01} and \texttt{Quantum Espresso} (QE) \cite{giannozzi:jpcm09}. The results obtained from these two codes
agree well. In QE, the pseudopotentials are generated by the included \texttt{ATOMIC} code. In the pseudopotential generation we used the fully relativistic PBESOL functional \cite{Wu:prb06}. Because QE can only accept norm-conserving pseudopotentials for finite spin-orbit coupling (SOC) when  Wannier projection is desired, we took all pseudopotentials to be norm-conserving.  The valence shell of Y ([Kr]4$d^15s^2$) includes the $4s, 4p, 4d, 5s, 5p$ orbitals, and the valence shell of 
Ir ([Xe]$4f^{14}5d^76s^2$) includes $5s, 5p, 5d, 6s$, and $6p$ orbitals. Including 
the semi-core states, that is, $4s, 4p$ in Y, and $5s, 5p$ in Ir improves the transferability of the pseudopotentials. The  valence states of O ([He]$2s^22p^4$) include the $2s$ and $2p$ states. The cutoff energy in  QE calculation was selected to be 150 Rydberg (Ry) for plane waves, and 600 Ry for the charge densities. The structure information of bulk Y$_2$Ir$_2$O$_7$ is from Ref.[\onlinecite{Taira:jop01}]. 

\section{II. Tight-binding fit to the bulk DFT calculation}

\begin{figure}[h]
\centering
\includegraphics[width=0.8\linewidth]{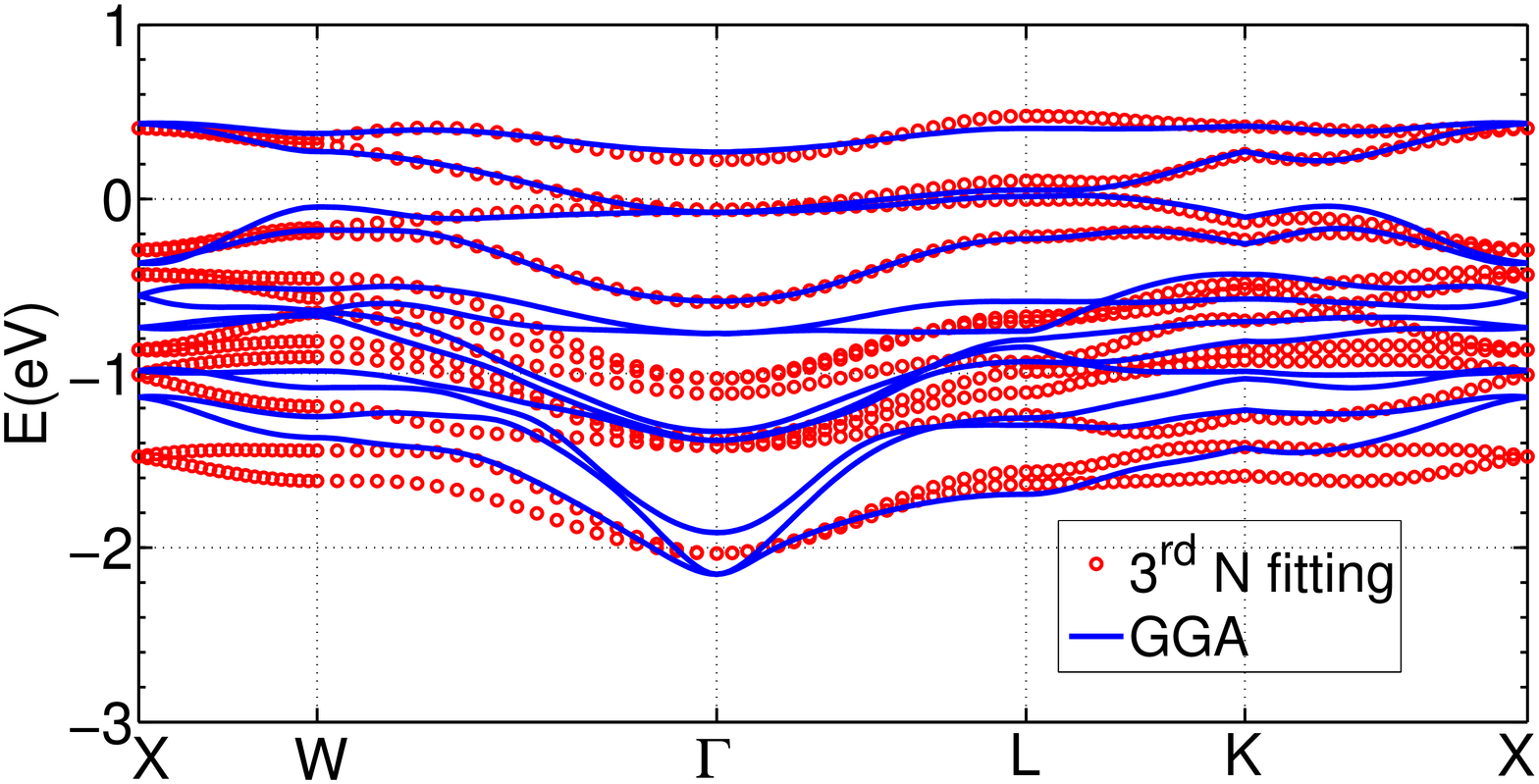}
\caption{Tight binding fit with up to the 3rd neighbor hopping compared
to a GGA calculation in \texttt{Wien2k}. \label{fig:TB_fit}}
\end{figure}
Because the tight-binding fitting method has been successfully applied to period four
transition metal oxides, such as LaNiO$_3$\cite{Rueggsup:prb12,Ruegg_Top:prb13}, we first tried 
to fit the DFT results for bulk Y$_2$Ir$_2$O$_7$ to a tight binding model that included both direct and indirect hopping between the local $t_{2g}$ orbitals of the Ir$^{4+}$ ions. The hopping matrix elements were generated using Slater-Koster (SK)\cite{SK:pr54} integrals.  As shown in Fig.\ref{fig:TB_fit} above, in order to obtain even a semi-quantitative fit for the $j=1/2$ bands (upper 4) one must include at least out to third-neighbor hoppings.  Even in this case, however, the lower $j=3/2$ bands are fit rather poorly.  The failure of the tight-binding fitting reveals the complexity of $5d$ orbits, and their much larger spatial extent compared to the $3d$ orbitals.  In order to obtain an satisfactory model, one must include more SK hopping terms in the tight-binding fitting, which is not well motivated for a practical model Hamiltonian. Instead, we turn to the Wannier states, making use of the \texttt{Wannier90} package\cite{Mostofi:cpc08}, which yields the fit shown in Fig. 1 of the main text. 

\section{III. Wannier projection of the bulk DFT calculation for Y$_2$I\MakeLowercase{r}$_2$O$_7$}

The spin-orbit coupling can be treated in two different ways based on Wannier projection. One route is to first perform a DFT calculation and  Wannier projection without SOC, and then add an onsite SOC ($\sim \lambda (-{\bf l)}\cdot{\bf s}$) term ``by hand" whose strength is determined through least-square fitting of the Wannier+SOC bands to the DFT bands with the SOC {\em turned on}\cite{Mostofi:cpc08, Yangsup:prl14}.  Alternatively, one can perform a DFT calculation with SOC turned on and then obtain a direct Wannier projection of the SOC coupled states \cite{Mostofi:cpc08}. 

In those Wannier projections, the initial basis is selected as the {\it local} $t_{2g}$ orbitals. Even though
the trigonal distortion causes a mixing between those $t_{2g}$ orbitals, the axis of the local $t_{2g}$ orbitals on each of the 4 sites in the unit cell can be assigned as\cite{Mostofi:cpc08} 
\begin{eqnarray}
x_1=(2/3, -1/3, 2/3), \nonumber \\
z_1=(-2/3, -2/3, 1/3); \nonumber \\
x_2=(2/3, -2/3, 1/3), \nonumber \\
z_2=(1/3, 2/3, 2/3); \nonumber \\
x_3=(1/3, 2/3, -2/3), \nonumber \\
z_3=(2/3, 1/3, 2/3); \nonumber \\
x_4=(1/3, -2/3, 2/3), \nonumber \\
z_4=(2/3, -1/3, -2/3).
\label{eq:coordf}
\end{eqnarray}
The coordinates in Eqs.(\ref{eq:coordf}) can be easily obtained from the rotational matrices $R^{(0)-(3)}$ in Ref.[\onlinecite{Pesinsup:np10}].
All the Wannier functions are centered on the iridium ions. The spin quantization axis for all 4 iridium ions in the unit cell is set to the the global $z$-axis.  To maximumly preserve the bulk crystal symmetry, the Num$\_$Iter is set to 0 in the Wannierization process. However, even if a finite Num$\_$Iter is used, the Wannier functions change only slightly, so the trial wave functions are quite good.  
 
\begin{figure}[h] 
\includegraphics[width=0.8\linewidth]{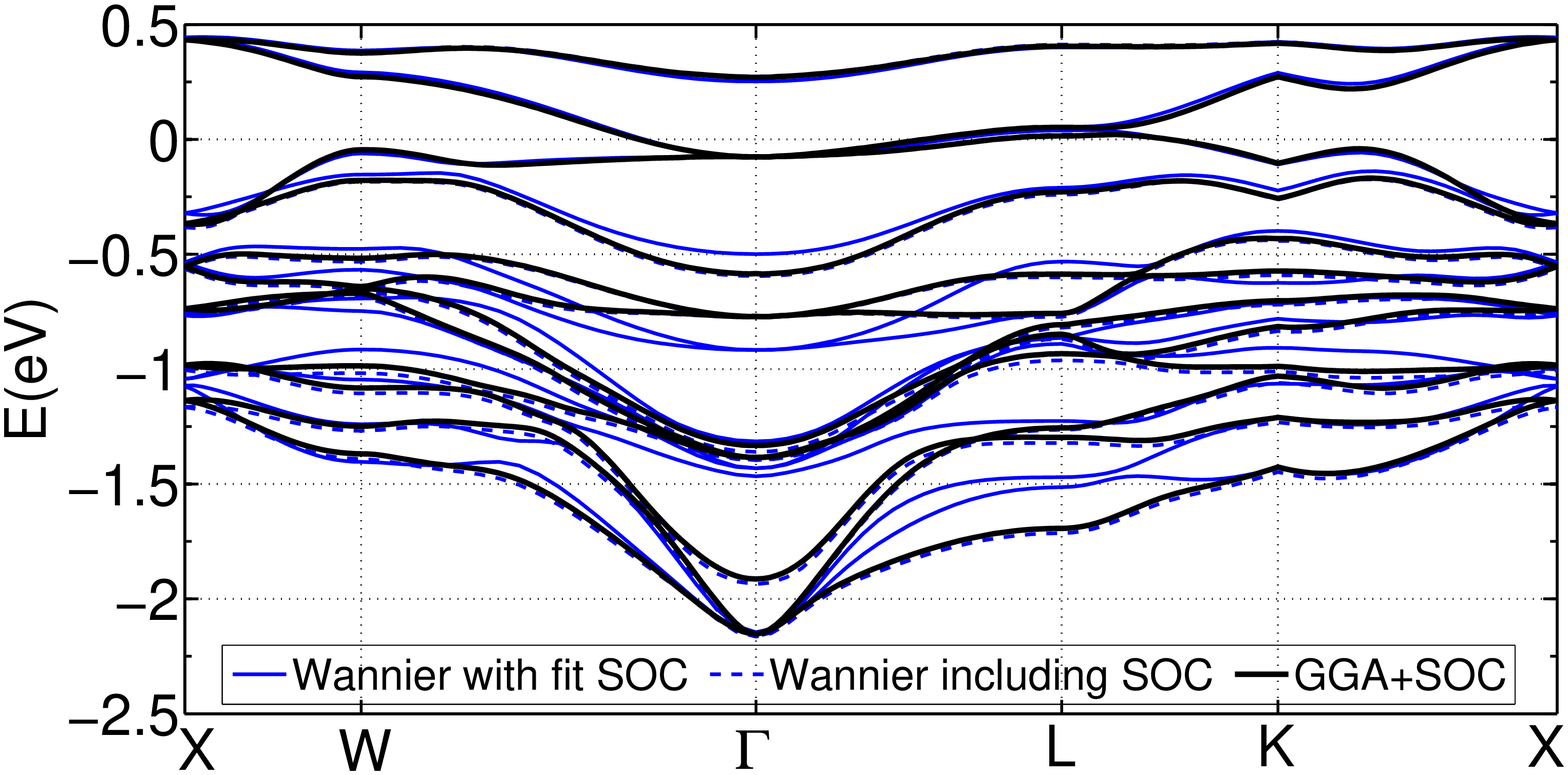}
\caption{A comparison of the two methods used to treat the spin-orbit coupling (SOC). The black curves are the band structure from the DFT calculation with SOC turned on for all atoms. The blue solid curves show the band structure obtained from the TB model.  The TB model was obtained from Wannier projection of the DFT calculation with SOC turned off supplemented with a  SOC added on ``by hand" and fit to the DFT band structure with the SOC turned on. The fit value of the SOC is 0.43eV. The dotted curves show the band structure resulting from the direct Wannier projection of the DFT calculation with SOC turned on. The DFT results were obtained through \texttt{Wien2k}. All the Wannier projections were done in \texttt{QE}+\texttt{Wannier90}. \label{fig:comp_fit}}
\end{figure}

When the SOC is fit ``by hand", we find a spin-orbit coupling strength of about 0.43 eV, corresponding roughly to the splitting of the $j=1/2$ and $j=3/2$ manifolds.  As shown in Fig.\ref{fig:comp_fit}, this leads to a reasonable fit to the fully relativistic (SOC on) DFT calculation. This approach should be able to capture the main character of the phases when interactions are further treated at the Hartree-Fock level. 
By comparison, a direct Wannier projection for the DFT calculation with SOC turned on captures the band feature more accurately (the blue dashed line in Fig.\ref{fig:comp_fit}) and will be more accurate in determining the phase diagrams.  (Also see Fig.1 of the main paper text. In the figure, the black line shows the GGA+SOC results obtained in \texttt{Wien2k}, while the blue dotted line shows the Wannier projection found through a post-processing of the DFT results obtained in \texttt{QE}.) 

From the analysis of the Wannier projection ones sees appreciable hopping between iridium ions that are as far as 3 to 4 FCC basis vectors from each other.  Our results agree with a previous study in Sr$_2$IrO$_4$, in which the hopping between Ir$^{4+}$ ions 1 nm (around 3 times the distance between our nearest neighbor hopping) away from each other still plays a role\cite{Kunes:cpc10}. 

\section{IV. Details of thin film DFT calculation for (Y$_2$I\MakeLowercase{r}$_2$O$_7$)$_n$/(Y$_2$H\MakeLowercase{f}$_2$O$_7$)$_m$}

In order to carry out our the thin film calculations, we construct the superlattice
with Y$_2$Ir$_2$O$_7$ sandwiched between the band insulator Y$_2$Hf$_2$O$_7$ \cite{rieken:ames11}.
We consider a superlattice (Y$_2$Ir$_2$O$_7$)$_n$/(Y$_2$Hf$_2$O$_7$)$_m$, where the sub-indexes $n$ and $m$ represent the numbers of Ir$^{4+}$ or Hf$^{4+}$ layers. To investigate the effect of strain, other substrates, such as Y$_2$Ti$_2$O$_7$ \cite{rieken:ames11} can also be used. 

The original FCC basis vectors for the bulk pyrochlore lattice are given by,
\begin{eqnarray*}
{\bf a}_1=(0, a/2, a/2),\\
{\bf a}_2=(a/2, 0, a/2),\\
{\bf a}_3=(a/2, a/2, 0),
\end{eqnarray*}
where $a$ is the lattice constant.

\begin{figure}[h] 
\centering
\includegraphics[width=0.35\linewidth]{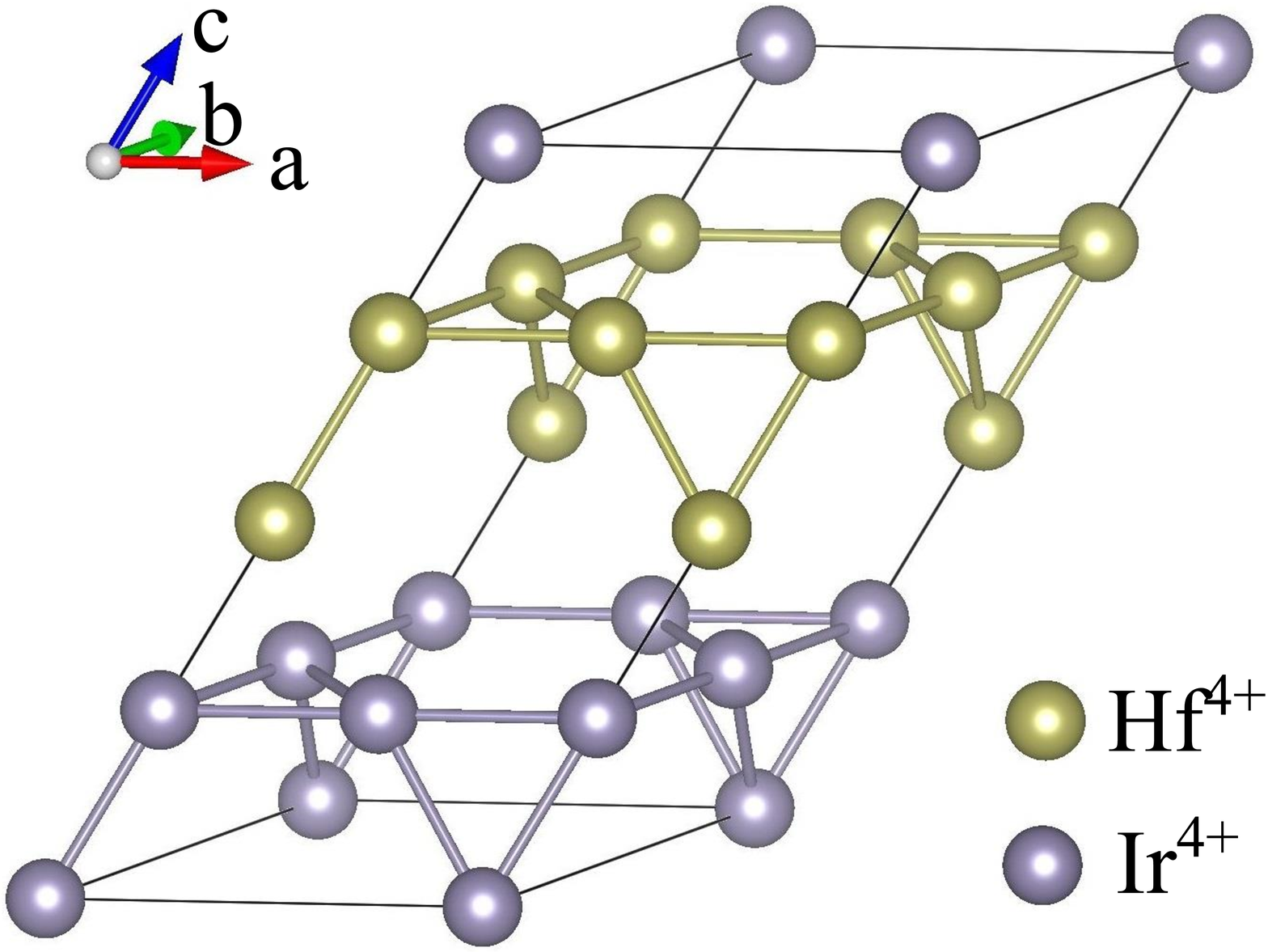}
\caption{The unrelaxed bilayer thin film. Only the Ir$^{4+}$ and Hf$^{4+}$ ions are displayed. \label{fig:bi-film}}
\end{figure}

To obtain the thin film structure along [111] direction, for the $n=2, m=2$ bilayer thin film, 
the basis vectors are selected as 
\begin{eqnarray*}
{\bf a}={\bf a}_3-{\bf a}_1 ,\\
{\bf b}={\bf a}_2-{\bf a}_1 ,\\
{\bf c}=-2*{\bf a}_1.
\end{eqnarray*}
Here ${\bf a}$ and ${\bf b}$ are in plane, while ${\bf c}$ contains a component out of the plane spanned by ${\bf a}$ and ${\bf b}$. The position of each ion can be expressed in terms of linear combinations of ${\bf a}, {\bf b}, {\bf c}$.  The size of the unit cell is been doubled compared to the bulk,  with Ir$^{4+}$ ions in two adjacent atomic layers substituted with Hf$^{4+}$ ions to obtain the unrelaxed bilayer structure shown in Fig.\ref{fig:bi-film}. As an initial guess, the lattice constant $a$ can be taken as the average of Y$_2$Hf$_2$O$_7$ and Y$_2$Ir$_2$O$_7$. The unrelaxed trilayer film structure can be obtained in a similar way. Some package such as \texttt{ASE Surface}\cite{Bahn:cse02} can also help to achieve this.

The determination of the lattice structure of the sandwhich is carried out in a fully relaxed scheme with scalar-relativistic\cite{Takeda:zpbcm78} pseudopotentials for both the bilayer superlattice $($Y$_2$Ir$_2$O$_{7})_2$/$($Y$_2$Hf$_2$O$_{7})_2$ and the trilayer superlattice $($Y$_2$Ir$_2$O$_{7})_3$/$($Y$_2$Hf$_2$O$_{7})_3$. We have verified that increasing the number of layers of Y$_2$Hf$_2$O$_{7}$ does not have a significant effect on the final lattice structures and the resulting band structures.  Once the lattice structure is determined, the band structures of the bilayer and trilayer Y$_2$Ir$_2$O$_{7}$ are calculated in a fully relativistic basis ({\it i.e.}, with the spin-orbit coupling turned on). The pseudopotential of Hf (with atomic configuration [Xe]$4f^{14}5d^26s^2$) is generated in the same way as the other elements, with the valence orbits as $5s, 5p, 5d, 6s$, and $6p$. The k-space is meshed by 7$\times$7$\times$1 with Monkhost-Pack Grid \cite{Monkhorst:prb76}.  The Wannier fits shown in Fig. 2 of the main text are based on the Wannier fits obtained for the {\em bulk} system, which are still fairly accurate for states near the Fermi energy.

\section{V. Thin film Hartree-Fock calculation}

The unrestricted Hartree-Fock calculation is carried out in the {\em full} $t_{2g}$ subspace of the iridium orbitals with Wannier states obtained from the bulk band structure.  The resulting fits are shown in Fig. 2 of the main text.  While the quality of the fits are less good than they are for the bulk (Fig. 1 of the main text), they are sufficient for producing reliable trends in terms of the phases expected as a function of the on-site interaction $U$.   Our thin film calculations start with 21 randomly generated coupling constants $m_{\alpha\beta}^i=\langle \hat{c}_{\alpha}^{i\dagger} \hat{c}_{\beta}^i\rangle$ per site, which are then iterated to convergence with a difference between $m_{\alpha\beta}^i$ input and output of less than 10$^{-8}$. The lattice size is selected by be 90 by 90 or 120 by 120. Around 
fifty sets of initial guesses are taken, and about two-thirds of them converge for most $U$ values.
The final convergent configuration with the minimum total energy is selected.

The magnetic moments are calculated in a way similar to the determination of the Lande-$g$ factor. The $g$ factor is obtained by 
\begin{equation}
g \langle {\bf j}^2\rangle=\langle (-g_l {\bf l}+ g_s {\bf s})\cdot {\bf j}\rangle,
\end{equation}
where ${\bf j}=-{\bf l}+{\bf s}$ is the total angular momentum in {\it local} coordinates. Here $-{\bf l}$ is the effective orbital angular momentum in $t_{2g}$ subspace obtained by $P_{t_{2g}}{\bf L}P_{t_{2g}}=-{\bf l}_{l=1}$, where $P_{t_{2g}}$ is the projection operation into the $t_{2g}$ subspace, and $g_l=1, g_s=2$.
Then the total magnetic momentum is determined by 
\begin{equation}
{\bf m}=g \mu_B \langle {\bf j}\rangle.
\end{equation}

Most of the magnetic moment originates in the $j=1/2$ subspace.   Similar to the situation reported in bulk Y$_2$Ir$_2$O$_{7}$ \cite{Wansup:prb11}, we find an energy difference (around 1-150 meV/site, depending on the distance from the magnetic phase transition) between states with different order.  All of our time-reversal symmetry broken solutions have non-zero net magnetization.  

To compute the $Z_2$ invariant in the non-magnetic insulators, we use the formulation of Ref.[\onlinecite{Fukui:jpsj07}]. We compute the Chern number in the magnetic insulating phases in a similar way\cite{Fukui:jpsj05}.   

\begin{figure}[h] 
\centering
\includegraphics[width=0.5\linewidth]{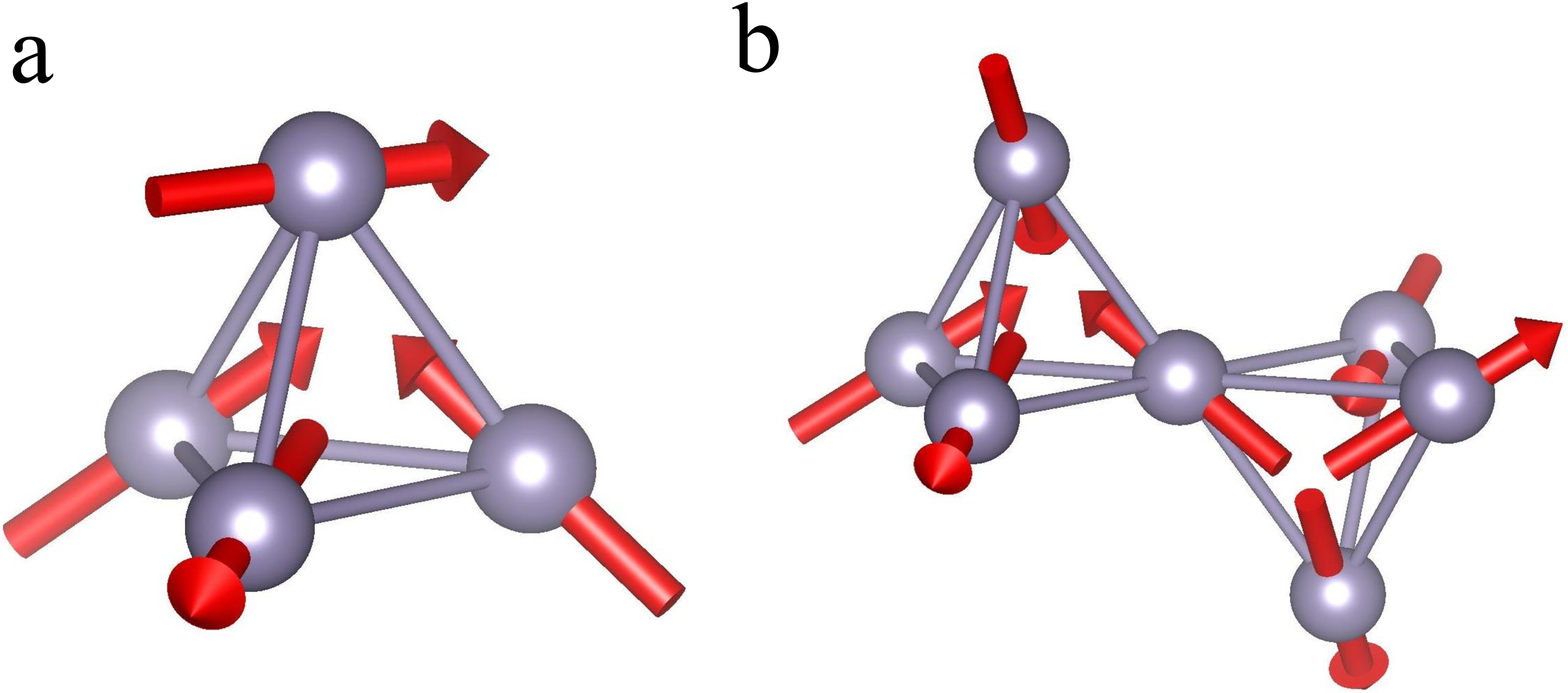}
\caption{The magnetic configurations at large $U$. (a) The magnetic configuration when $U=1.30$eV in bilayer thin film. (b) The magnetic configuration when $U=1.20$eV in TKT thin film. Both of them are in the magnetic insulating phase. In both (a) and (b), all the three in-plane ions in one unitcell possess magnetic moments with the same magnitude.\label{fig:orientations}}
\end{figure}

The magnetic metallic phase is close to the nonmagnetic to magnetic phase transition point, so the magnetic moments may have different configurations with very close energy. However, for a magnetic
insulating phase with large $U$, the magnetic configuration is rather stable, and has the pattern shown in Fig.\ref{fig:orientations}.


\begin{thebibliography}{30}%
	\makeatletter
	\providecommand \@ifxundefined [1]{%
		\@ifx{#1\undefined}
	}%
	\providecommand \@ifnum [1]{%
		\ifnum #1\expandafter \@firstoftwo
		\else \expandafter \@secondoftwo
		\fi
	}%
	\providecommand \@ifx [1]{%
		\ifx #1\expandafter \@firstoftwo
		\else \expandafter \@secondoftwo
		\fi
	}%
	\providecommand \natexlab [1]{#1}%
	\providecommand \enquote  [1]{``#1''}%
	\providecommand \bibnamefont  [1]{#1}%
	\providecommand \bibfnamefont [1]{#1}%
	\providecommand \citenamefont [1]{#1}%
	\providecommand \href@noop [0]{\@secondoftwo}%
	\providecommand \href [0]{\begingroup \@sanitize@url \@href}%
	\providecommand \@href[1]{\@@startlink{#1}\@@href}%
	\providecommand \@@href[1]{\endgroup#1\@@endlink}%
	\providecommand \@sanitize@url [0]{\catcode `\\12\catcode `\$12\catcode
		`\&12\catcode `\#12\catcode `\^12\catcode `\_12\catcode `\%12\relax}%
	\providecommand \@@startlink[1]{}%
	\providecommand \@@endlink[0]{}%
	\providecommand \url  [0]{\begingroup\@sanitize@url \@url }%
	\providecommand \@url [1]{\endgroup\@href {#1}{\urlprefix }}%
	\providecommand \urlprefix  [0]{URL }%
	\providecommand \Eprint [0]{\href }%
	\providecommand \doibase [0]{http://dx.doi.org/}%
	\providecommand \selectlanguage [0]{\@gobble}%
	\providecommand \bibinfo  [0]{\@secondoftwo}%
	\providecommand \bibfield  [0]{\@secondoftwo}%
	\providecommand \translation [1]{[#1]}%
	\providecommand \BibitemOpen [0]{}%
	\providecommand \bibitemStop [0]{}%
	\providecommand \bibitemNoStop [0]{.\EOS\space}%
	\providecommand \EOS [0]{\spacefactor3000\relax}%
	\providecommand \BibitemShut  [1]{\csname bibitem#1\endcsname}%
	\let\auto@bib@innerbib\@empty
	\bibitem [{\citenamefont {Moore}(2010)}]{Moore:nat10}%
	\BibitemOpen
	\bibfield  {author} {\bibinfo {author} {\bibfnamefont {J.~E.}\ \bibnamefont
			{Moore}},\ }\href {\doibase 10.1038/nature08916} {\bibfield  {journal}
		{\bibinfo  {journal} {Nature}\ }\textbf {\bibinfo {volume} {464}},\ \bibinfo
		{pages} {194} (\bibinfo {year} {2010})}\BibitemShut {NoStop}%
	\bibitem [{\citenamefont {Hasan}\ and\ \citenamefont
		{Kane}(2010)}]{Hasan:rmp10}%
	\BibitemOpen
	\bibfield  {author} {\bibinfo {author} {\bibfnamefont {M.~Z.}\ \bibnamefont
			{Hasan}}\ and\ \bibinfo {author} {\bibfnamefont {C.~L.}\ \bibnamefont
			{Kane}},\ }\href {\doibase 10.1103/RevModPhys.82.3045} {\bibfield  {journal}
		{\bibinfo  {journal} {Rev. Mod. Phys.}\ }\textbf {\bibinfo {volume} {82}},\
		\bibinfo {pages} {3045} (\bibinfo {year} {2010})}\BibitemShut {NoStop}%
	\bibitem [{\citenamefont {Qi}\ and\ \citenamefont {Zhang}(2011)}]{Qi:rmp11}%
	\BibitemOpen
	\bibfield  {author} {\bibinfo {author} {\bibfnamefont {X.-L.}\ \bibnamefont
			{Qi}}\ and\ \bibinfo {author} {\bibfnamefont {S.-C.}\ \bibnamefont {Zhang}},\
	}\href {\doibase 10.1103/RevModPhys.83.1057} {\bibfield  {journal} {\bibinfo
		{journal} {Rev. Mod. Phys.}\ }\textbf {\bibinfo {volume} {83}},\ \bibinfo
	{pages} {1057} (\bibinfo {year} {2011})}\BibitemShut {NoStop}%
\bibitem [{\citenamefont {Ando}(2013)}]{Ando:jpsj13}%
\BibitemOpen
\bibfield  {author} {\bibinfo {author} {\bibfnamefont {Y.}~\bibnamefont
		{Ando}},\ }\href@noop {} {\bibfield  {journal} {\bibinfo  {journal} {J. Phys.
			Soc. Jpn.}\ }\textbf {\bibinfo {volume} {82}},\ \bibinfo {pages} {102001}
	(\bibinfo {year} {2013})}\BibitemShut {NoStop}%
\bibitem [{\citenamefont {K\"onig}\ \emph {et~al.}(2007)\citenamefont
	{K\"onig}, \citenamefont {Wiedmann}, \citenamefont {Brune}, \citenamefont
	{Roth}, \citenamefont {Buhmann}, \citenamefont {Molenkamp}, \citenamefont
	{Qi},\ and\ \citenamefont {Zhang}}]{Konig:sci07}%
\BibitemOpen
\bibfield  {author} {\bibinfo {author} {\bibfnamefont {M.}~\bibnamefont
		{K\"onig}}, \bibinfo {author} {\bibfnamefont {S.}~\bibnamefont {Wiedmann}},
	\bibinfo {author} {\bibfnamefont {C.}~\bibnamefont {Brune}}, \bibinfo
	{author} {\bibfnamefont {A.}~\bibnamefont {Roth}}, \bibinfo {author}
	{\bibfnamefont {H.}~\bibnamefont {Buhmann}}, \bibinfo {author} {\bibfnamefont
		{L.}~\bibnamefont {Molenkamp}}, \bibinfo {author} {\bibfnamefont {X.-L.}\
		\bibnamefont {Qi}}, \ and\ \bibinfo {author} {\bibfnamefont {S.-C.}\
		\bibnamefont {Zhang}},\ }\href {\doibase 10.1126/science.1148047} {\bibfield
	{journal} {\bibinfo  {journal} {Science}\ }\textbf {\bibinfo {volume}
		{318}},\ \bibinfo {pages} {766} (\bibinfo {year} {2007})}\BibitemShut
{NoStop}%
\bibitem [{\citenamefont {Knez}\ \emph {et~al.}(2011)\citenamefont {Knez},
	\citenamefont {Du},\ and\ \citenamefont {Sullivan}}]{Knez:prl11}%
\BibitemOpen
\bibfield  {author} {\bibinfo {author} {\bibfnamefont {I.}~\bibnamefont
		{Knez}}, \bibinfo {author} {\bibfnamefont {R.-R.}\ \bibnamefont {Du}}, \ and\
	\bibinfo {author} {\bibfnamefont {G.}~\bibnamefont {Sullivan}},\ }\href@noop
{} {\bibfield  {journal} {\bibinfo  {journal} {Phys. Rev. Lett.}\ }\textbf
	{\bibinfo {volume} {107}},\ \bibinfo {pages} {136603} (\bibinfo {year}
	{2011})}\BibitemShut {NoStop}%
\bibitem [{\citenamefont {Chang}\ \emph {et~al.}(2013)\citenamefont {Chang},
	\citenamefont {Zhang}, \citenamefont {Feng}, \citenamefont {Shen},
	\citenamefont {Zhang}, \citenamefont {Guo}, \citenamefont {Li}, \citenamefont
	{Ou}, \citenamefont {Wei}, \citenamefont {Wang}, \citenamefont {Ji},
	\citenamefont {Feng}, \citenamefont {Shuaihua}, \citenamefont {Chen},
	\citenamefont {Jia}, \citenamefont {Dai}, \citenamefont {Fang}, \citenamefont
	{Zhang}, \citenamefont {He}, \citenamefont {Wang}, \citenamefont {Lu},
	\citenamefont {Ma},\ and\ \citenamefont {Xue}}]{Chang:sci13}%
\BibitemOpen
\bibfield  {author} {\bibinfo {author} {\bibfnamefont {C.-Z.}\ \bibnamefont
		{Chang}}, \bibinfo {author} {\bibfnamefont {J.}~\bibnamefont {Zhang}},
	\bibinfo {author} {\bibfnamefont {X.}~\bibnamefont {Feng}}, \bibinfo {author}
	{\bibfnamefont {J.}~\bibnamefont {Shen}}, \bibinfo {author} {\bibfnamefont
		{Z.}~\bibnamefont {Zhang}}, \bibinfo {author} {\bibfnamefont
		{M.}~\bibnamefont {Guo}}, \bibinfo {author} {\bibfnamefont {K.}~\bibnamefont
		{Li}}, \bibinfo {author} {\bibfnamefont {Y.}~\bibnamefont {Ou}}, \bibinfo
	{author} {\bibfnamefont {P.}~\bibnamefont {Wei}}, \bibinfo {author}
	{\bibfnamefont {L.-L.}\ \bibnamefont {Wang}}, \bibinfo {author}
	{\bibfnamefont {Z.-Q.}\ \bibnamefont {Ji}}, \bibinfo {author} {\bibfnamefont
		{Y.}~\bibnamefont {Feng}}, \bibinfo {author} {\bibnamefont {Shuaihua}},
	\bibinfo {author} {\bibfnamefont {X.}~\bibnamefont {Chen}}, \bibinfo {author}
	{\bibfnamefont {J.}~\bibnamefont {Jia}}, \bibinfo {author} {\bibfnamefont
		{X.}~\bibnamefont {Dai}}, \bibinfo {author} {\bibfnamefont {Z.}~\bibnamefont
		{Fang}}, \bibinfo {author} {\bibfnamefont {S.-C.}\ \bibnamefont {Zhang}},
	\bibinfo {author} {\bibfnamefont {K.}~\bibnamefont {He}}, \bibinfo {author}
	{\bibfnamefont {Y.}~\bibnamefont {Wang}}, \bibinfo {author} {\bibfnamefont
		{L.}~\bibnamefont {Lu}}, \bibinfo {author} {\bibfnamefont {X.-C.}\
		\bibnamefont {Ma}}, \ and\ \bibinfo {author} {\bibfnamefont {Q.-K.}\
		\bibnamefont {Xue}},\ }\href {\doibase 10.1126/science.1234414} {\bibfield
	{journal} {\bibinfo  {journal} {Science}\ }\textbf {\bibinfo {volume}
		{167}},\ \bibinfo {pages} {167} (\bibinfo {year} {2013})}\BibitemShut
{NoStop}%
\bibitem [{\citenamefont {Bergman}\ and\ \citenamefont
	{Refael}(2010)}]{Bergman:prb10}%
\BibitemOpen
\bibfield  {author} {\bibinfo {author} {\bibfnamefont {D.~L.}\ \bibnamefont
		{Bergman}}\ and\ \bibinfo {author} {\bibfnamefont {G.}~\bibnamefont
		{Refael}},\ }\href {\doibase 10.1103/PhysRevB.82.195417} {\bibfield
	{journal} {\bibinfo  {journal} {Phys. Rev. B}\ }\textbf {\bibinfo {volume}
		{82}},\ \bibinfo {pages} {195417} (\bibinfo {year} {2010})}\BibitemShut
{NoStop}%
\bibitem [{\citenamefont {Pan}\ \emph {et~al.}(2014)\citenamefont {Pan},
	\citenamefont {Li}, \citenamefont {Qiao}, \citenamefont {Liu}, \citenamefont
	{Yao},\ and\ \citenamefont {Yang}}]{Pan_top_met_14}%
\BibitemOpen
\bibfield  {author} {\bibinfo {author} {\bibfnamefont {H.}~\bibnamefont
		{Pan}}, \bibinfo {author} {\bibfnamefont {X.}~\bibnamefont {Li}}, \bibinfo
	{author} {\bibfnamefont {Z.}~\bibnamefont {Qiao}}, \bibinfo {author}
	{\bibfnamefont {C.-C.}\ \bibnamefont {Liu}}, \bibinfo {author} {\bibfnamefont
		{Y.}~\bibnamefont {Yao}}, \ and\ \bibinfo {author} {\bibfnamefont {S.~A.}\
		\bibnamefont {Yang}},\ }\href@noop {} {\enquote {\bibinfo {title}
		{Topological metallic states in spin-orbit coupled bilayer systems},}\ }
(\bibinfo {year} {2014}),\ \bibinfo {note} {arXiv:1402.6502}\BibitemShut
{NoStop}%
\bibitem [{\citenamefont {Hwang}\ \emph {et~al.}(2012)\citenamefont {Hwang},
	\citenamefont {Iwasa}, \citenamefont {Kawasaki}, \citenamefont {Keimer},
	\citenamefont {Nagaosa},\ and\ \citenamefont {Tokura}}]{Hwang:nm12}%
\BibitemOpen
\bibfield  {author} {\bibinfo {author} {\bibfnamefont {H.~Y.}\ \bibnamefont
		{Hwang}}, \bibinfo {author} {\bibfnamefont {Y.}~\bibnamefont {Iwasa}},
	\bibinfo {author} {\bibfnamefont {M.}~\bibnamefont {Kawasaki}}, \bibinfo
	{author} {\bibfnamefont {B.}~\bibnamefont {Keimer}}, \bibinfo {author}
	{\bibfnamefont {N.}~\bibnamefont {Nagaosa}}, \ and\ \bibinfo {author}
	{\bibfnamefont {Y.}~\bibnamefont {Tokura}},\ }\href {\doibase
	10.1038/nmat3223} {\bibfield  {journal} {\bibinfo  {journal} {Nat. Mat.}\
	}\textbf {\bibinfo {volume} {11}},\ \bibinfo {pages} {103} (\bibinfo {year}
	{2012})}\BibitemShut {NoStop}%
\bibitem [{\citenamefont {Witczak-Krempa}\ \emph {et~al.}(2014)\citenamefont
	{Witczak-Krempa}, \citenamefont {Chen}, \citenamefont {Kim},\ and\
	\citenamefont {Balents}}]{Krempa:arcm14}%
\BibitemOpen
\bibfield  {author} {\bibinfo {author} {\bibfnamefont {W.}~\bibnamefont
		{Witczak-Krempa}}, \bibinfo {author} {\bibfnamefont {G.}~\bibnamefont
		{Chen}}, \bibinfo {author} {\bibfnamefont {Y.~B.}\ \bibnamefont {Kim}}, \
	and\ \bibinfo {author} {\bibfnamefont {L.}~\bibnamefont {Balents}},\
}\href@noop {} {\bibfield  {journal} {\bibinfo  {journal} {Ann. Rev. Cond.
		Matt. Phys.}\ }\textbf {\bibinfo {volume} {5}},\ \bibinfo {pages} {57}
(\bibinfo {year} {2014})}\BibitemShut {NoStop}%
\bibitem [{\citenamefont {Pesin}\ and\ \citenamefont
	{Balents}(2010)}]{Pesin:np10}%
\BibitemOpen
\bibfield  {author} {\bibinfo {author} {\bibfnamefont {D.}~\bibnamefont
		{Pesin}}\ and\ \bibinfo {author} {\bibfnamefont {L.}~\bibnamefont
		{Balents}},\ }\href {\doibase 10.1038/nphys1606} {\bibfield  {journal}
	{\bibinfo  {journal} {Nat. Phys.}\ }\textbf {\bibinfo {volume} {6}},\
	\bibinfo {pages} {376} (\bibinfo {year} {2010})}\BibitemShut {NoStop}%
\bibitem [{\citenamefont {Kargarian}\ \emph {et~al.}(2011)\citenamefont
	{Kargarian}, \citenamefont {Wen},\ and\ \citenamefont
	{Fiete}}]{Kargarian:prb11}%
\BibitemOpen
\bibfield  {author} {\bibinfo {author} {\bibfnamefont {M.}~\bibnamefont
		{Kargarian}}, \bibinfo {author} {\bibfnamefont {J.}~\bibnamefont {Wen}}, \
	and\ \bibinfo {author} {\bibfnamefont {G.~A.}\ \bibnamefont {Fiete}},\ }\href
{\doibase 10.1103/PhysRevB.83.165112} {\bibfield  {journal} {\bibinfo
		{journal} {Phys. Rev. B}\ }\textbf {\bibinfo {volume} {83}},\ \bibinfo
	{pages} {165112} (\bibinfo {year} {2011})}\BibitemShut {NoStop}%
\bibitem [{\citenamefont {Kargarian}\ and\ \citenamefont
	{Fiete}(2013)}]{Kargarian:prl13}%
\BibitemOpen
\bibfield  {author} {\bibinfo {author} {\bibfnamefont {M.}~\bibnamefont
		{Kargarian}}\ and\ \bibinfo {author} {\bibfnamefont {G.~A.}\ \bibnamefont
		{Fiete}},\ }\href {\doibase 10.1103/PhysRevLett.110.156403} {\bibfield
	{journal} {\bibinfo  {journal} {Phys. Rev. Lett.}\ }\textbf {\bibinfo
		{volume} {110}},\ \bibinfo {pages} {156403} (\bibinfo {year}
	{2013})}\BibitemShut {NoStop}%
\bibitem [{\citenamefont {Maciejko}\ \emph {et~al.}(2014)\citenamefont
	{Maciejko}, \citenamefont {Chua},\ and\ \citenamefont
	{Fiete}}]{Maciejko:prl14}%
\BibitemOpen
\bibfield  {author} {\bibinfo {author} {\bibfnamefont {J.}~\bibnamefont
		{Maciejko}}, \bibinfo {author} {\bibfnamefont {V.}~\bibnamefont {Chua}}, \
	and\ \bibinfo {author} {\bibfnamefont {G.~A.}\ \bibnamefont {Fiete}},\
}\href@noop {} {\bibfield  {journal} {\bibinfo  {journal} {Phys. Rev. Lett.}\
}\textbf {\bibinfo {volume} {112}},\ \bibinfo {pages} {016404} (\bibinfo
{year} {2014})}\BibitemShut {NoStop}%
\bibitem [{\citenamefont {Go}\ \emph {et~al.}(2012)\citenamefont {Go},
	\citenamefont {Witczak-Krempa}, \citenamefont {Jeon}, \citenamefont {Park},\
	and\ \citenamefont {Kim}}]{Go:prl12}%
\BibitemOpen
\bibfield  {author} {\bibinfo {author} {\bibfnamefont {A.}~\bibnamefont
		{Go}}, \bibinfo {author} {\bibfnamefont {W.}~\bibnamefont {Witczak-Krempa}},
	\bibinfo {author} {\bibfnamefont {G.~S.}\ \bibnamefont {Jeon}}, \bibinfo
	{author} {\bibfnamefont {K.}~\bibnamefont {Park}}, \ and\ \bibinfo {author}
	{\bibfnamefont {Y.~B.}\ \bibnamefont {Kim}},\ }\href {\doibase
	10.1103/PhysRevLett.109.066401} {\bibfield  {journal} {\bibinfo  {journal}
		{Phys. Rev. Lett.}\ }\textbf {\bibinfo {volume} {109}},\ \bibinfo {pages}
	{066401} (\bibinfo {year} {2012})}\BibitemShut {NoStop}%
\bibitem [{\citenamefont {Wan}\ \emph {et~al.}(2011)\citenamefont {Wan},
	\citenamefont {Turner}, \citenamefont {Vishwanath},\ and\ \citenamefont
	{Savrasov}}]{Wan:prb11}%
\BibitemOpen
\bibfield  {author} {\bibinfo {author} {\bibfnamefont {X.}~\bibnamefont
		{Wan}}, \bibinfo {author} {\bibfnamefont {A.~M.}\ \bibnamefont {Turner}},
	\bibinfo {author} {\bibfnamefont {A.}~\bibnamefont {Vishwanath}}, \ and\
	\bibinfo {author} {\bibfnamefont {S.~Y.}\ \bibnamefont {Savrasov}},\ }\href
{\doibase 10.1103/PhysRevB.83.205101} {\bibfield  {journal} {\bibinfo
		{journal} {Phys. Rev. B}\ }\textbf {\bibinfo {volume} {83}},\ \bibinfo
	{pages} {205101} (\bibinfo {year} {2011})}\BibitemShut {NoStop}%
\bibitem [{\citenamefont {R\"uegg}\ and\ \citenamefont
	{Fiete}(2011)}]{Ruegg11_2}%
\BibitemOpen
\bibfield  {author} {\bibinfo {author} {\bibfnamefont {A.}~\bibnamefont
		{R\"uegg}}\ and\ \bibinfo {author} {\bibfnamefont {G.~A.}\ \bibnamefont
		{Fiete}},\ }\href {\doibase 10.1103/PhysRevB.84.201103} {\bibfield  {journal}
	{\bibinfo  {journal} {Phys. Rev. B}\ }\textbf {\bibinfo {volume} {84}},\
	\bibinfo {pages} {201103} (\bibinfo {year} {2011})}\BibitemShut {NoStop}%
\bibitem [{\citenamefont {Yang}\ \emph {et~al.}(2011)\citenamefont {Yang},
	\citenamefont {Zhu}, \citenamefont {Xiao}, \citenamefont {Okamoto},
	\citenamefont {Wang},\ and\ \citenamefont {Ran}}]{Yang:prb11a}%
\BibitemOpen
\bibfield  {author} {\bibinfo {author} {\bibfnamefont {K.-Y.}\ \bibnamefont
		{Yang}}, \bibinfo {author} {\bibfnamefont {W.}~\bibnamefont {Zhu}}, \bibinfo
	{author} {\bibfnamefont {D.}~\bibnamefont {Xiao}}, \bibinfo {author}
	{\bibfnamefont {S.}~\bibnamefont {Okamoto}}, \bibinfo {author} {\bibfnamefont
		{Z.}~\bibnamefont {Wang}}, \ and\ \bibinfo {author} {\bibfnamefont
		{Y.}~\bibnamefont {Ran}},\ }\href {\doibase 10.1103/PhysRevB.84.201104}
{\bibfield  {journal} {\bibinfo  {journal} {Phys. Rev. B}\ }\textbf {\bibinfo
		{volume} {84}},\ \bibinfo {pages} {201104} (\bibinfo {year}
	{2011})}\BibitemShut {NoStop}%
\bibitem [{\citenamefont {R\"uegg}\ \emph {et~al.}(2012)\citenamefont
	{R\"uegg}, \citenamefont {Mitra}, \citenamefont {Demkov},\ and\ \citenamefont
	{Fiete}}]{Ruegg:prb12}%
\BibitemOpen
\bibfield  {author} {\bibinfo {author} {\bibfnamefont {A.}~\bibnamefont
		{R\"uegg}}, \bibinfo {author} {\bibfnamefont {C.}~\bibnamefont {Mitra}},
	\bibinfo {author} {\bibfnamefont {A.~A.}\ \bibnamefont {Demkov}}, \ and\
	\bibinfo {author} {\bibfnamefont {G.~A.}\ \bibnamefont {Fiete}},\ }\href
{\doibase 10.1103/PhysRevB.85.245131} {\bibfield  {journal} {\bibinfo
		{journal} {Phys. Rev. B}\ }\textbf {\bibinfo {volume} {85}},\ \bibinfo
	{pages} {245131} (\bibinfo {year} {2012})}\BibitemShut {NoStop}%
\bibitem [{\citenamefont {Xiao}\ \emph {et~al.}(2011)\citenamefont {Xiao},
	\citenamefont {Zhu}, \citenamefont {Ran}, \citenamefont {Nagaosa},\ and\
	\citenamefont {Okamoto}}]{Xiao:nc11}%
\BibitemOpen
\bibfield  {author} {\bibinfo {author} {\bibfnamefont {D.}~\bibnamefont
		{Xiao}}, \bibinfo {author} {\bibfnamefont {W.}~\bibnamefont {Zhu}}, \bibinfo
	{author} {\bibfnamefont {Y.}~\bibnamefont {Ran}}, \bibinfo {author}
	{\bibfnamefont {N.}~\bibnamefont {Nagaosa}}, \ and\ \bibinfo {author}
	{\bibfnamefont {S.}~\bibnamefont {Okamoto}},\ }\href {\doibase
	10.1038/ncomms1602} {\bibfield  {journal} {\bibinfo  {journal} {Nat. Comm.}\
	}\textbf {\bibinfo {volume} {2}},\ \bibinfo {pages} {596} (\bibinfo {year}
	{2011})}\BibitemShut {NoStop}%
\bibitem [{\citenamefont {Okamoto}\ \emph {et~al.}(2014)\citenamefont
	{Okamoto}, \citenamefont {Zhu}, \citenamefont {Nomura}, \citenamefont
	{Arita}, \citenamefont {Xiao},\ and\ \citenamefont
	{Nagaosa}}]{Okamoto:prb14}%
\BibitemOpen
\bibfield  {author} {\bibinfo {author} {\bibfnamefont {S.}~\bibnamefont
		{Okamoto}}, \bibinfo {author} {\bibfnamefont {W.}~\bibnamefont {Zhu}},
	\bibinfo {author} {\bibfnamefont {Y.}~\bibnamefont {Nomura}}, \bibinfo
	{author} {\bibfnamefont {R.}~\bibnamefont {Arita}}, \bibinfo {author}
	{\bibfnamefont {D.}~\bibnamefont {Xiao}}, \ and\ \bibinfo {author}
	{\bibfnamefont {N.}~\bibnamefont {Nagaosa}},\ }\href {\doibase
	10.1103/PhysRevB.89.195121} {\bibfield  {journal} {\bibinfo  {journal} {Phys.
			Rev. B}\ }\textbf {\bibinfo {volume} {89}},\ \bibinfo {pages} {195121}
	(\bibinfo {year} {2014})}\BibitemShut {NoStop}%
\bibitem [{\citenamefont {Yang}\ and\ \citenamefont
	{Nagaosa}(2014)}]{Yang:prl14}%
\BibitemOpen
\bibfield  {author} {\bibinfo {author} {\bibfnamefont {B.-J.}\ \bibnamefont
		{Yang}}\ and\ \bibinfo {author} {\bibfnamefont {N.}~\bibnamefont {Nagaosa}},\
}\href {\doibase 10.1103/PhysRevLett.112.246402} {\bibfield  {journal}
{\bibinfo  {journal} {Phys. Rev. Lett.}\ }\textbf {\bibinfo {volume} {112}},\
\bibinfo {pages} {246402} (\bibinfo {year} {2014})}\BibitemShut {NoStop}%
\bibitem [{\citenamefont {Okamoto}(2013)}]{Okamoto:prl13}%
\BibitemOpen
\bibfield  {author} {\bibinfo {author} {\bibfnamefont {S.}~\bibnamefont
		{Okamoto}},\ }\href {\doibase 10.1103/PhysRevLett.110.066403} {\bibfield
	{journal} {\bibinfo  {journal} {Phys. Rev. Lett.}\ }\textbf {\bibinfo
		{volume} {110}},\ \bibinfo {pages} {066403} (\bibinfo {year}
	{2013})}\BibitemShut {NoStop}%
\bibitem [{\citenamefont {Doennig}\ \emph {et~al.}(2014)\citenamefont
	{Doennig}, \citenamefont {Pickett},\ and\ \citenamefont
	{Pentcheva}}]{Doennig:prb14}%
\BibitemOpen
\bibfield  {author} {\bibinfo {author} {\bibfnamefont {D.}~\bibnamefont
		{Doennig}}, \bibinfo {author} {\bibfnamefont {W.~E.}\ \bibnamefont
		{Pickett}}, \ and\ \bibinfo {author} {\bibfnamefont {R.}~\bibnamefont
		{Pentcheva}},\ }\href {\doibase 10.1103/PhysRevB.89.121110} {\bibfield
	{journal} {\bibinfo  {journal} {Phys. Rev. B}\ }\textbf {\bibinfo {volume}
		{89}},\ \bibinfo {pages} {121110} (\bibinfo {year} {2014})}\BibitemShut
{NoStop}%
\bibitem [{\citenamefont {Lado}\ \emph {et~al.}(2013)\citenamefont {Lado},
	\citenamefont {Pardo},\ and\ \citenamefont {Baldomir}}]{Lado:prb13}%
\BibitemOpen
\bibfield  {author} {\bibinfo {author} {\bibfnamefont {J.~L.}\ \bibnamefont
		{Lado}}, \bibinfo {author} {\bibfnamefont {V.}~\bibnamefont {Pardo}}, \ and\
	\bibinfo {author} {\bibfnamefont {D.}~\bibnamefont {Baldomir}},\ }\href
{\doibase 10.1103/PhysRevB.88.155119} {\bibfield  {journal} {\bibinfo
		{journal} {Phys. Rev. B}\ }\textbf {\bibinfo {volume} {88}},\ \bibinfo
	{pages} {155119} (\bibinfo {year} {2013})}\BibitemShut {NoStop}%
\bibitem [{\citenamefont {Liang}\ \emph {et~al.}(2013)\citenamefont {Liang},
	\citenamefont {Wu},\ and\ \citenamefont {Hu}}]{Liang:njp13}%
\BibitemOpen
\bibfield  {author} {\bibinfo {author} {\bibfnamefont {Q.-F.}\ \bibnamefont
		{Liang}}, \bibinfo {author} {\bibfnamefont {L.-H.}\ \bibnamefont {Wu}}, \
	and\ \bibinfo {author} {\bibfnamefont {X.}~\bibnamefont {Hu}},\ }\href
{\doibase 10.1088/1367-2630/15/6/063031} {\bibfield  {journal} {\bibinfo
		{journal} {New Journal of Physics}\ }\textbf {\bibinfo {volume} {15}},\
	\bibinfo {pages} {063031} (\bibinfo {year} {2013})}\BibitemShut {NoStop}%
\bibitem [{\citenamefont {Wang}\ \emph {et~al.}(2014)\citenamefont {Wang},
	\citenamefont {Wang}, \citenamefont {Fang},\ and\ \citenamefont
	{Dai}}]{Wang14}%
\BibitemOpen
\bibfield  {author} {\bibinfo {author} {\bibfnamefont {Y.}~\bibnamefont
		{Wang}}, \bibinfo {author} {\bibfnamefont {Z.}~\bibnamefont {Wang}}, \bibinfo
	{author} {\bibfnamefont {Z.}~\bibnamefont {Fang}}, \ and\ \bibinfo {author}
	{\bibfnamefont {X.}~\bibnamefont {Dai}},\ }\href@noop {} {\enquote {\bibinfo
		{title} {Interaction-induced quantum anomalous hall phase in (111) bilayer of
			lacoo$_3$},}\ } (\bibinfo {year} {2014}),\ \bibinfo {note}
{arXiv:1409.6797}\BibitemShut {NoStop}%
\bibitem [{\citenamefont {Hu}\ \emph {et~al.}(2012)\citenamefont {Hu},
	\citenamefont {R\"uegg},\ and\ \citenamefont {Fiete}}]{Hu:prb12}%
\BibitemOpen
\bibfield  {author} {\bibinfo {author} {\bibfnamefont {X.}~\bibnamefont
		{Hu}}, \bibinfo {author} {\bibfnamefont {A.}~\bibnamefont {R\"uegg}}, \ and\
	\bibinfo {author} {\bibfnamefont {G.~A.}\ \bibnamefont {Fiete}},\ }\href
{\doibase 10.1103/PhysRevB.86.235141} {\bibfield  {journal} {\bibinfo
		{journal} {Phys. Rev. B}\ }\textbf {\bibinfo {volume} {86}},\ \bibinfo
	{pages} {235141} (\bibinfo {year} {2012})}\BibitemShut {NoStop}%
\bibitem [{\citenamefont {Zhang}\ \emph {et~al.}(2013)\citenamefont {Zhang},
	\citenamefont {Haule},\ and\ \citenamefont {Vanderbilt}}]{Zhang_Huale:prl13}%
\BibitemOpen
\bibfield  {author} {\bibinfo {author} {\bibfnamefont {H.}~\bibnamefont
		{Zhang}}, \bibinfo {author} {\bibfnamefont {K.}~\bibnamefont {Haule}}, \ and\
	\bibinfo {author} {\bibfnamefont {D.}~\bibnamefont {Vanderbilt}},\ }\href
{\doibase 10.1103/PhysRevLett.111.246402} {\bibfield  {journal} {\bibinfo
		{journal} {Phys. Rev. Lett.}\ }\textbf {\bibinfo {volume} {111}},\ \bibinfo
	{pages} {246402} (\bibinfo {year} {2013})}\BibitemShut {NoStop}%
\end{thebibliography}

\begin{thebibliography}{18}%
\makeatletter
\providecommand \@ifxundefined [1]{%
 \@ifx{#1\undefined}
}%
\providecommand \@ifnum [1]{%
 \ifnum #1\expandafter \@firstoftwo
 \else \expandafter \@secondoftwo
 \fi
}%
\providecommand \@ifx [1]{%
 \ifx #1\expandafter \@firstoftwo
 \else \expandafter \@secondoftwo
 \fi
}%
\providecommand \natexlab [1]{#1}%
\providecommand \enquote  [1]{``#1''}%
\providecommand \bibnamefont  [1]{#1}%
\providecommand \bibfnamefont [1]{#1}%
\providecommand \citenamefont [1]{#1}%
\providecommand \href@noop [0]{\@secondoftwo}%
\providecommand \href [0]{\begingroup \@sanitize@url \@href}%
\providecommand \@href[1]{\@@startlink{#1}\@@href}%
\providecommand \@@href[1]{\endgroup#1\@@endlink}%
\providecommand \@sanitize@url [0]{\catcode `\\12\catcode `\$12\catcode
  `\&12\catcode `\#12\catcode `\^12\catcode `\_12\catcode `\%12\relax}%
\providecommand \@@startlink[1]{}%
\providecommand \@@endlink[0]{}%
\providecommand \url  [0]{\begingroup\@sanitize@url \@url }%
\providecommand \@url [1]{\endgroup\@href {#1}{\urlprefix }}%
\providecommand \urlprefix  [0]{URL }%
\providecommand \Eprint [0]{\href }%
\providecommand \doibase [0]{http://dx.doi.org/}%
\providecommand \selectlanguage [0]{\@gobble}%
\providecommand \bibinfo  [0]{\@secondoftwo}%
\providecommand \bibfield  [0]{\@secondoftwo}%
\providecommand \translation [1]{[#1]}%
\providecommand \BibitemOpen [0]{}%
\providecommand \bibitemStop [0]{}%
\providecommand \bibitemNoStop [0]{.\EOS\space}%
\providecommand \EOS [0]{\spacefactor3000\relax}%
\providecommand \BibitemShut  [1]{\csname bibitem#1\endcsname}%
\let\auto@bib@innerbib\@empty
\bibitem [{\citenamefont {Blaha}\ \emph {et~al.}(2001)\citenamefont {Blaha},
  \citenamefont {Schwarz}, \citenamefont {Madsen}, \citenamefont {Kvasnicka},\
  and\ \citenamefont {Luitz}}]{Blaha:tu01}%
  \BibitemOpen
  \bibfield  {author} {\bibinfo {author} {\bibfnamefont {P.}~\bibnamefont
  {Blaha}}, \bibinfo {author} {\bibfnamefont {K.}~\bibnamefont {Schwarz}},
  \bibinfo {author} {\bibfnamefont {G.~K.~H.}\ \bibnamefont {Madsen}}, \bibinfo
  {author} {\bibfnamefont {D.}~\bibnamefont {Kvasnicka}}, \ and\ \bibinfo
  {author} {\bibfnamefont {J.}~\bibnamefont {Luitz}},\ }\href@noop {} {\emph
  {\bibinfo {title} {{WIEN2K}, {A}n {A}ugmented {P}lane {W}ave + {L}ocal
  {O}rbitals {P}rogram for {C}alculating {C}rystal {P}roperties}}}\ (\bibinfo
  {publisher} {{K}arlheinz Schwarz, Techn. Universit\"{a}t Wien, Austria},\
  \bibinfo {year} {2001})\BibitemShut {NoStop}%
\bibitem [{\citenamefont {Giannozzi}\ \emph {et~al.}(2009)\citenamefont
  {Giannozzi}, \citenamefont {Baroni}, \citenamefont {Bonini} \emph
  {et~al.}}]{giannozzi:jpcm09}%
  \BibitemOpen
  \bibfield  {author} {\bibinfo {author} {\bibfnamefont {P.}~\bibnamefont
  {Giannozzi}}, \bibinfo {author} {\bibfnamefont {S.}~\bibnamefont {Baroni}},
  \bibinfo {author} {\bibfnamefont {N.}~\bibnamefont {Bonini}},  \emph
  {et~al.},\ }\href@noop {} {\bibfield  {journal} {\bibinfo  {journal} {J.
  Phys.: Condens. Matter}\ }\textbf {\bibinfo {volume} {21}},\ \bibinfo {pages}
  {395502} (\bibinfo {year} {2009})}\BibitemShut {NoStop}%
\bibitem [{\citenamefont {Wu}\ and\ \citenamefont {Cohen}(2006)}]{Wu:prb06}%
  \BibitemOpen
  \bibfield  {author} {\bibinfo {author} {\bibfnamefont {Z.}~\bibnamefont
  {Wu}}\ and\ \bibinfo {author} {\bibfnamefont {R.~E.}\ \bibnamefont {Cohen}},\
  }\href {\doibase 10.1103/PhysRevB.73.235116} {\bibfield  {journal} {\bibinfo
  {journal} {Phys. Rev. B}\ }\textbf {\bibinfo {volume} {73}},\ \bibinfo
  {pages} {235116} (\bibinfo {year} {2006})}\BibitemShut {NoStop}%
\bibitem [{\citenamefont {Taira}\ \emph {et~al.}(2001)\citenamefont {Taira},
  \citenamefont {Wakeshima},\ and\ \citenamefont {Hinatsu}}]{Taira:jop01}%
  \BibitemOpen
  \bibfield  {author} {\bibinfo {author} {\bibfnamefont {N.}~\bibnamefont
  {Taira}}, \bibinfo {author} {\bibfnamefont {M.}~\bibnamefont {Wakeshima}}, \
  and\ \bibinfo {author} {\bibfnamefont {Y.}~\bibnamefont {Hinatsu}},\
  }\href@noop {} {\bibfield  {journal} {\bibinfo  {journal} {J. Phys.: Condens.
  Matter}\ }\textbf {\bibinfo {volume} {13}},\ \bibinfo {pages} {5527}
  (\bibinfo {year} {2001})}\BibitemShut {NoStop}%
\bibitem [{\citenamefont {R\"uegg}\ \emph {et~al.}(2012)\citenamefont
  {R\"uegg}, \citenamefont {Mitra}, \citenamefont {Demkov},\ and\ \citenamefont
  {Fiete}}]{Rueggsup:prb12}%
  \BibitemOpen
  \bibfield  {author} {\bibinfo {author} {\bibfnamefont {A.}~\bibnamefont
  {R\"uegg}}, \bibinfo {author} {\bibfnamefont {C.}~\bibnamefont {Mitra}},
  \bibinfo {author} {\bibfnamefont {A.~A.}\ \bibnamefont {Demkov}}, \ and\
  \bibinfo {author} {\bibfnamefont {G.~A.}\ \bibnamefont {Fiete}},\ }\href
  {\doibase 10.1103/PhysRevB.85.245131} {\bibfield  {journal} {\bibinfo
  {journal} {Phys. Rev. B}\ }\textbf {\bibinfo {volume} {85}},\ \bibinfo
  {pages} {245131} (\bibinfo {year} {2012})}\BibitemShut {NoStop}%
\bibitem [{\citenamefont {R\"uegg}\ \emph {et~al.}(2013)\citenamefont
  {R\"uegg}, \citenamefont {Mitra}, \citenamefont {Demkov},\ and\ \citenamefont
  {Fiete}}]{Ruegg_Top:prb13}%
  \BibitemOpen
  \bibfield  {author} {\bibinfo {author} {\bibfnamefont {A.}~\bibnamefont
  {R\"uegg}}, \bibinfo {author} {\bibfnamefont {C.}~\bibnamefont {Mitra}},
  \bibinfo {author} {\bibfnamefont {A.~A.}\ \bibnamefont {Demkov}}, \ and\
  \bibinfo {author} {\bibfnamefont {G.~A.}\ \bibnamefont {Fiete}},\ }\href
  {\doibase 10.1103/PhysRevB.88.115146} {\bibfield  {journal} {\bibinfo
  {journal} {Phys. Rev. B}\ }\textbf {\bibinfo {volume} {88}},\ \bibinfo
  {pages} {115146} (\bibinfo {year} {2013})}\BibitemShut {NoStop}%
\bibitem [{\citenamefont {Slater}\ and\ \citenamefont
  {Koster}(1954)}]{SK:pr54}%
  \BibitemOpen
  \bibfield  {author} {\bibinfo {author} {\bibfnamefont {J.~C.}\ \bibnamefont
  {Slater}}\ and\ \bibinfo {author} {\bibfnamefont {G.~F.}\ \bibnamefont
  {Koster}},\ }\href {\doibase 10.1103/PhysRev.94.1498} {\bibfield  {journal}
  {\bibinfo  {journal} {Phys. Rev.}\ }\textbf {\bibinfo {volume} {94}},\
  \bibinfo {pages} {1498} (\bibinfo {year} {1954})}\BibitemShut {NoStop}%
\bibitem [{\citenamefont {Mostofi}\ \emph {et~al.}(2008)\citenamefont
  {Mostofi}, \citenamefont {Yates}, \citenamefont {Lee}, \citenamefont {Souza},
  \citenamefont {Vanderbilt},\ and\ \citenamefont {Marzari}}]{Mostofi:cpc08}%
  \BibitemOpen
  \bibfield  {author} {\bibinfo {author} {\bibfnamefont {A.~A.}\ \bibnamefont
  {Mostofi}}, \bibinfo {author} {\bibfnamefont {J.~R.}\ \bibnamefont {Yates}},
  \bibinfo {author} {\bibfnamefont {Y.-S.}\ \bibnamefont {Lee}}, \bibinfo
  {author} {\bibfnamefont {I.}~\bibnamefont {Souza}}, \bibinfo {author}
  {\bibfnamefont {D.}~\bibnamefont {Vanderbilt}}, \ and\ \bibinfo {author}
  {\bibfnamefont {N.}~\bibnamefont {Marzari}},\ }\href@noop {} {\bibfield
  {journal} {\bibinfo  {journal} {Comput. Phys. Commun.}\ }\textbf {\bibinfo
  {volume} {178}},\ \bibinfo {pages} {685} (\bibinfo {year}
  {2008})}\BibitemShut {NoStop}%
\bibitem [{\citenamefont {Yang}\ and\ \citenamefont
  {Nagaosa}(2014)}]{Yangsup:prl14}%
  \BibitemOpen
  \bibfield  {author} {\bibinfo {author} {\bibfnamefont {B.-J.}\ \bibnamefont
  {Yang}}\ and\ \bibinfo {author} {\bibfnamefont {N.}~\bibnamefont {Nagaosa}},\
  }\href {\doibase 10.1103/PhysRevLett.112.246402} {\bibfield  {journal}
  {\bibinfo  {journal} {Phys. Rev. Lett.}\ }\textbf {\bibinfo {volume} {112}},\
  \bibinfo {pages} {246402} (\bibinfo {year} {2014})}\BibitemShut {NoStop}%
\bibitem [{\citenamefont {Pesin}\ and\ \citenamefont
  {Balents}(2010)}]{Pesinsup:np10}%
  \BibitemOpen
  \bibfield  {author} {\bibinfo {author} {\bibfnamefont {D.}~\bibnamefont
  {Pesin}}\ and\ \bibinfo {author} {\bibfnamefont {L.}~\bibnamefont
  {Balents}},\ }\href {\doibase 10.1038/nphys1606} {\bibfield  {journal}
  {\bibinfo  {journal} {Nat. Phys.}\ }\textbf {\bibinfo {volume} {6}},\
  \bibinfo {pages} {376} (\bibinfo {year} {2010})}\BibitemShut {NoStop}%
\bibitem [{\citenamefont {Kune?}\ \emph {et~al.}(2010)\citenamefont {Kune?},
  \citenamefont {Arita}, \citenamefont {Wissgott}, \citenamefont {Toschi},
  \citenamefont {Ikeda},\ and\ \citenamefont {Held}}]{Kunes:cpc10}%
  \BibitemOpen
  \bibfield  {author} {\bibinfo {author} {\bibfnamefont {J.}~\bibnamefont
  {Kune}}, \bibinfo {author} {\bibfnamefont {R.}~\bibnamefont {Arita}},
  \bibinfo {author} {\bibfnamefont {P.}~\bibnamefont {Wissgott}}, \bibinfo
  {author} {\bibfnamefont {A.}~\bibnamefont {Toschi}}, \bibinfo {author}
  {\bibfnamefont {H.}~\bibnamefont {Ikeda}}, \ and\ \bibinfo {author}
  {\bibfnamefont {K.}~\bibnamefont {Held}},\ }\href {\doibase
  http://dx.doi.org/10.1016/j.cpc.2010.08.005} {\bibfield  {journal} {\bibinfo
  {journal} {Comput. Phys. Commun.}\ }\textbf {\bibinfo {volume} {181}},\
  \bibinfo {pages} {1888 } (\bibinfo {year} {2010})}\BibitemShut {NoStop}%
\bibitem [{\citenamefont {Rieken}\ \emph {et~al.}(2011)\citenamefont {Rieken},
  \citenamefont {Anderson},\ and\ \citenamefont {Kramer}}]{rieken:ames11}%
  \BibitemOpen
  \bibfield  {author} {\bibinfo {author} {\bibfnamefont {J.}~\bibnamefont
  {Rieken}}, \bibinfo {author} {\bibfnamefont {I.}~\bibnamefont {Anderson}}, \
  and\ \bibinfo {author} {\bibfnamefont {M.}~\bibnamefont {Kramer}},\
  }\href@noop {} {\emph {\bibinfo {title} {Innovative Powder Processing of
  Oxide Dispersion Strengthened ODS Ferritic Stainless Steels}}},\ \bibinfo
  {type} {Tech. Rep.}\ (\bibinfo  {institution} {Ames Laboratory (AMES), Ames,
  IA (United States)},\ \bibinfo {year} {2011})\BibitemShut {NoStop}%
\bibitem [{\citenamefont {Bahn}\ and\ \citenamefont
  {Jacobsen}(2002)}]{Bahn:cse02}%
  \BibitemOpen
  \bibfield  {author} {\bibinfo {author} {\bibfnamefont {S.~R.}\ \bibnamefont
  {Bahn}}\ and\ \bibinfo {author} {\bibfnamefont {K.~W.}\ \bibnamefont
  {Jacobsen}},\ }\href {\doibase 10.1109/5992.998641} {\bibfield  {journal}
  {\bibinfo  {journal} {Comput. Sci. Eng.}\ }\textbf {\bibinfo {volume} {4}},\
  \bibinfo {pages} {56} (\bibinfo {year} {2002})}\BibitemShut {NoStop}%
\bibitem [{\citenamefont {Takeda}(1978)}]{Takeda:zpbcm78}%
  \BibitemOpen
  \bibfield  {author} {\bibinfo {author} {\bibfnamefont {T.}~\bibnamefont
  {Takeda}},\ }\href {\doibase 10.1007/BF01322185} {\bibfield  {journal}
  {\bibinfo  {journal} {Z. Phys. B Con. Mat.}\ }\textbf {\bibinfo {volume}
  {32}},\ \bibinfo {pages} {43} (\bibinfo {year} {1978})}\BibitemShut {NoStop}%
\bibitem [{\citenamefont {Monkhorst}\ and\ \citenamefont
  {Pack}(1976)}]{Monkhorst:prb76}%
  \BibitemOpen
  \bibfield  {author} {\bibinfo {author} {\bibfnamefont {H.~J.}\ \bibnamefont
  {Monkhorst}}\ and\ \bibinfo {author} {\bibfnamefont {J.~D.}\ \bibnamefont
  {Pack}},\ }\href {\doibase 10.1103/PhysRevB.13.5188} {\bibfield  {journal}
  {\bibinfo  {journal} {Phys. Rev. B}\ }\textbf {\bibinfo {volume} {13}},\
  \bibinfo {pages} {5188} (\bibinfo {year} {1976})}\BibitemShut {NoStop}%
\bibitem [{\citenamefont {Wan}\ \emph {et~al.}(2011)\citenamefont {Wan},
  \citenamefont {Turner}, \citenamefont {Vishwanath},\ and\ \citenamefont
  {Savrasov}}]{Wansup:prb11}%
  \BibitemOpen
  \bibfield  {author} {\bibinfo {author} {\bibfnamefont {X.}~\bibnamefont
  {Wan}}, \bibinfo {author} {\bibfnamefont {A.~M.}\ \bibnamefont {Turner}},
  \bibinfo {author} {\bibfnamefont {A.}~\bibnamefont {Vishwanath}}, \ and\
  \bibinfo {author} {\bibfnamefont {S.~Y.}\ \bibnamefont {Savrasov}},\ }\href
  {\doibase 10.1103/PhysRevB.83.205101} {\bibfield  {journal} {\bibinfo
  {journal} {Phys. Rev. B}\ }\textbf {\bibinfo {volume} {83}},\ \bibinfo
  {pages} {205101} (\bibinfo {year} {2011})}\BibitemShut {NoStop}%
\bibitem [{\citenamefont {Fukui}\ and\ \citenamefont
  {Hatsugai}(2007)}]{Fukui:jpsj07}%
  \BibitemOpen
  \bibfield  {author} {\bibinfo {author} {\bibfnamefont {T.}~\bibnamefont
  {Fukui}}\ and\ \bibinfo {author} {\bibfnamefont {Y.}~\bibnamefont
  {Hatsugai}},\ }\href {\doibase 10.1143/JPSJ.76.053702} {\bibfield  {journal}
  {\bibinfo  {journal} {J. Phys. Soc. Jap.}\ }\textbf {\bibinfo {volume}
  {76}},\ \bibinfo {pages} {053702} (\bibinfo {year} {2007})}\BibitemShut
  {NoStop}%
\bibitem [{\citenamefont {Fukui}\ \emph {et~al.}(2005)\citenamefont {Fukui},
  \citenamefont {Hatsugai},\ and\ \citenamefont {Suzuki}}]{Fukui:jpsj05}%
  \BibitemOpen
  \bibfield  {author} {\bibinfo {author} {\bibfnamefont {T.}~\bibnamefont
  {Fukui}}, \bibinfo {author} {\bibfnamefont {Y.}~\bibnamefont {Hatsugai}}, \
  and\ \bibinfo {author} {\bibfnamefont {H.}~\bibnamefont {Suzuki}},\
  }\href@noop {} {\bibfield  {journal} {\bibinfo  {journal} {J. Phys. Soc.
  Jap.}\ }\textbf {\bibinfo {volume} {74}},\ \bibinfo {pages} {1674} (\bibinfo
  {year} {2005})}\BibitemShut {NoStop}%
\end{thebibliography}
\end{document}